\documentclass[twocolumn]{aastex62}

\pdfoutput=1

\newcommand{\kms}{km~s$^{-1}$}

\newcommand{\NaI}{Na~{\sc i}}

\newcommand{\FeII}{Fe~{\sc ii}}

\newcommand{\CoII}{Co~{\sc ii}}

\newcommand{\NiII}{Ni~{\sc ii}}

\newcommand{\Nifs}{$^{56}$Ni}

\newcommand{\sBV}{s$_{BV}$}

\newcommand{\ab}{$\sim$}

\newcommand{\ved}{$v_{edge}$}

\graphicspath{{./}{figures/}}

\shorttitle{Sibling Type Ia Supernovae}
\shortauthors{Burns, Ashall, Contreras, Brown, et al.}

\begin{document}

\title{SN~2013aa and SN~2017cbv: Two Sibling Type~Ia Supernovae in the spiral galaxy NGC~5643\footnote{This paper includes data gathered with the 6.5-meter Magellan Telescopes located at Las Campanas Observatory, Chile.}}

\correspondingauthor{Christopher Burns}
\email{cburns@carnegiescience.edu}
\author[0000-0003-4625-6629]{Christopher~R.~Burns}
\affiliation{Observatories of the Carnegie Institution for Science, 813 Santa Barbara St, Pasadena, CA, 91101, USA}
\author[0000-0002-5221-7557]{Chris~Ashall}
\affiliation{Department of Physics, Florida State University, Tallahassee, FL 32306, USA}
\author[0000-0001-6293-9062]{Carlos~Contreras}
\affiliation{Carnegie Observatories, Las Campanas Observatory, Casilla 601, Chile}
\author[0000-0001-6272-5507]{Peter Brown}
\affiliation{George P. and Cynthia Woods Mitchell Institute for Fundamental Physics and Astronomy, Texas A\&M University, Department of Physics and Astronomy, College Station, TX, 77843, USA}
\author[0000-0002-5571-1833]{Maximilian~Stritzinger}
\affiliation{Department of Physics and Astronomy, Aarhus University, Ny Munkegade 120, DK-8000 Aarhus C, Denmark}
\author[0000-0003-2734-0796]{M.~M.~Phillips}
\affiliation{Carnegie Observatories, Las Campanas Observatory, Casilla 601, Chile}
\author{Ricardo Flores}
\affiliation{Department of Physics \& Astronomy, San Francisco State University,
1600 Holloway Avenue, San Francisco, CA, 94132, USA}
\author{Nicholas~B.~Suntzeff}
\affiliation{George P. and Cynthia Woods Mitchell Institute for Fundamental Physics and Astronomy, Texas A\&M University, Department of Physics and Astronomy, College Station, TX, 77843, USA}
\author[0000-0003-1039-2928]{Eric~Y.~Hsiao}
\affiliation{Department of Physics, Florida State University, Tallahassee, FL 32306, USA}
\author[0000-0002-9413-4186]{Syed Uddin}
\affiliation{Observatories of the Carnegie Institution for Science, 813 Santa Barbara St, Pasadena, CA, 91101, USA}
\author{Joshua~D.~Simon}
\affiliation{Observatories of the Carnegie Institution for Science, 813 Santa Barbara St, Pasadena, CA, 91101, USA}
\author[0000-0002-6650-694X]{Kevin~Krisciunas}
\affiliation{George P. and Cynthia Woods Mitchell Institute for Fundamental Physics and Astronomy, Texas A\&M University, Department of Physics and Astronomy, College Station, TX, 77843, USA}
\author[0000-0002-3829-9920]{Abdo Campillay}
\affiliation{Departamento de F\'isica y Astronom\'ia, Universidad de La Serena, Av. Cisternas 1200 Norte, La Serena, Chile}
\author{Ryan~J.~Foley}
\affiliation{Department of Astronomy and Astrophysics, University of California, Santa Cruz, CA 95064, USA}
\author[0000-0003-3431-9135]{Wendy~L.~Freedman}
\affiliation{Department of Astronomy \& Astrophysics, University of Chicago, 5640 South Ellis Avenue, Chicago, IL 60637, USA}
\author[0000-0002-1296-6887]{Llu\'is Galbany}
\affiliation{Departamento de F\'isica Te\'orica y del Cosmos, Universidad de Granada, E-18071 Granada, Spain}
\author{Consuelo~Gonz\'{a}lez}
\affiliation{Carnegie Observatories, Las Campanas Observatory, Casilla 601, Chile}
\author[0000-0002-4338-6586]{Peter~Hoeflich}
\affiliation{Department of Physics, Florida State University, Tallahassee, FL 32306, USA}
\author{S. Holmbo}
\affiliation{Department of Physics and Astronomy, Aarhus University, Ny Munkegade 120, DK-8000 Aarhus C, Denmark}
\author[0000-0002-5740-7747]{Charles D. Kilpatrick}
\affiliation{Department of Astronomy and Astrophysics, University of California, Santa Cruz, CA 95064, USA}
\author[0000-0002-1966-3942]{Robert P. Kirshner}
\affiliation{Gordon and Betty Moore Foundation, 1661 Page Mill Road, Palo Alto, CA 94304, USA}
\affiliation{Harvard-Smithsonian Center for Astrophysics, 60 Garden Street, Cambridge, MA 02138, USA}
\author{Nidia~Morrell}
\affiliation{Carnegie Observatories, Las Campanas Observatory, Casilla 601, La Serena, Chile}
\author{Nahir Mu\~noz-Elgueta}
\affiliation{Departamento de F\'isica y Astronom\'ia, Universidad de La Serena, Av. Cisternas 1200 Norte, La Serena, Chile}
\author{Anthony~L.~Piro}
\affiliation{Observatories of the Carnegie Institution for Science, 813 Santa Barbara St, Pasadena, CA, 91101, USA}
\author[0000-0002-7559-315X]{C\'esar Rojas-Bravo}
\affiliation{Department of Astronomy and Astrophysics, University of California, Santa Cruz, CA 95064, USA}
\author[0000-0003-4102-380X]{David~Sand}
\affiliation{Department of Astronomy/Steward Observatory, 933 North Cherry Avenue, Rm. N204, Tucson, AZ 85721-0065, USA}
\author[0000-0003-4329-3299]{Jaime~Vargas-Gonz\'{a}lez}
\affiliation{Centre for Astrophysics Research, School of Physics, Astronomy and Mathematics, University of Hertfordshire, College Lane, Hatfield AL10 9AB, UK}
\author{Natalie Ulloa}
\affiliation{Departamento de F\'isica y Astronom\'ia, Universidad de La Serena, Av. Cisternas 1200 Norte, La Serena, Chile}
\author{Jorge Anais Vilchez}
\affiliation{Centro de Astronom\'{i}a (CITEVA), Universidad the Antofagasta, Avenida Angamos 601, Antofagasta, Chile}

\submitjournal{\apj}
\accepted{April 26, 2020}

\begin{abstract}
We present photometric and spectroscopic observations of SN~2013aa and
SN~2017cbv, two nearly identical type~Ia supernovae (SNe~Ia) in the
host galaxy NGC~5643. The optical photometry has been obtained using
the same telescope and instruments used by the \textit{Carnegie Supernova Project}.
This eliminates most instrumental systematics and provides light curves in a stable and
well-understood photometric system. 
Having the same host galaxy also eliminates systematics due to 
distance and peculiar velocity, providing
an opportunity to directly test the relative precision of SNe~Ia as
standard candles.
The two SNe have nearly identical decline rates,
negligible reddening,
and remarkably similar spectra and, at a distance of $\sim 20$ Mpc,
are ideal as potential calibrators for the absolute distance using primary indicators
such as Cepheid variables. We discuss to what extent these two SNe can
be considered twins and compare them with other supernova ``siblings"
in the literature and their likely progenitor scenarios. Using 
12 galaxies that hosted 2 or more SNe~Ia, we find that when using SNe~Ia, 
 and after accounting for all sources of observational error,
one gets consistency in distance to 3\%. 
\end{abstract}


\section{Introduction} \label{sec:intro}

Through the application of empirical-based calibration techniques
\citep{Phillips:1993,Hamuy:1996,Phillips:1999}, type~Ia supernovae (SNe~Ia)
have served as robust extragalactic distance indicators.  Early work focused on
the intrinsic scatter with respect to a first or second order fit of absolute
magnitudes at maximum light vs. decline rate at optical wavelengths
\citep{Hamuy:1996, Riess:1999}.  The scatter in this {\em luminosity-decline
rate relation} ranged between nearly 0.2 mag in the $U$ band, to 0.15 mag in
the  $I$ band. Later work leveraging the near-infrared (NIR), where dimming due
to dust is an order of magnitude lower \citep{Fitzpatrick:1999}, showed similar
dispersions \citep{Krisciunas:2004,Wood-Vasey:2008,Folatelli:2010,Kattner:2012}
with possibly the lowest scatter at $H$ band \citep{Mandel:2009}. 

Working in the spectral domain, \citet{Fakhouri:2015} introduced the notion
of SN~Ia twins, which are SNe~Ia that have similar spectral features and therefore are
expected to have similar progenitor systems and explosion scenarios. 
They showed that sub-dividing
the sample into bins of like ``twinness" results in dispersions in distances to the SNe of $\sim 0.08$
mag. Then again, \citet{Foley:2018b} demonstrated that two such twins (SN~2011by
and SN~2011fe) appear to differ in intrinsic luminosity by 
$\Delta M_V = 0.335 \pm 0.069$ mag.

Most of these analyses, however,
are based on low red-shift ($z \lesssim 0.1)$ samples, which are prone
to extra variance due to the peculiar velocities and bulk
flows of their host galaxies relative to cosmic expansion.
Studying two (or more) SNe~Ia hosted by the same galaxy, 
which we shall call ``siblings" \citep{Brown:2015}, 
offers the possibility of comparing their inferred
distances without this extra uncertainty, allowing a better
estimate of the errors involved.
Studying supernova siblings also mitigates any extra systematics that may be 
correlated with  host-galaxy properties 
\citep{Sullivan:2010,Kelly:2010,Lampeitl:2010}.

The first study of SN~Ia siblings was by \citet{Hamuy:1991}, who 
considered
NGC~1316 (Fornax A), which hosted two normal SNe~Ia: SN~1980N and SN~1981D. 
Using un-corrected peak magnitudes of these SNe, they found the inferred distances
differed by $\sim 0.1$ mag. Two and a half decades later, NGC~1316 produced 
two more SNe~Ia, one normal (SN~2006dd) as well as one fast-decliner (SN~2006mr).
Over the same time span,
the methods for standardizing SN~Ia distances had significantly improved
\citep{Pskovskii:1984,Phillips:1992,Riess:1996}.
\citet{Stritzinger:2011} compared all
four siblings using these updated methods.
They found a dispersion of $4\%-8\%$ in distance, however much of that was
likely due to differences in photometric systems, some of which were 
difficult to
characterize. They also found a larger discrepancy ($\sim 25\%$) with respect
to the fast-decliner (SN~2006mr), though this was later found to be
due to $\Delta m_{15}(B)$ being a poor measure of the decline
rate for the fastest-declining SNe~Ia. Using the color-stretch parameter
$s_{BV}$ \citep{Burns:2018} instead reduced the discrepancy to less than the 
measurement errors.

\citet{Gall18} compared the distances from two ``transitional" SNe~Ia 
\citep{Pastorello07,Hsiao:2015,Ashall16a}: SN~2007on and SN~2011iv 
hosted by NGC~1404, another Fornax cluster
member. In this case, the photometric systems were identical, yet the discrepancy
in distances was $\sim 14\%$ in the optical and $\sim 9\%$ in the
NIR. It was argued that the observed discrepancy must be due to physical 
differences in the progenitors of both systems. 
More specifically, the central densities of the progenitor  white dwarfs (WDs) were hypothesized
to differ \citep{Gall18,Hoeflich17,Ashall18}.

In this paper, we consider two SNe~Ia
hosted by the spiral galaxy NGC~5643: SN~2013aa and SN~2017cbv. Both have 
extensive optical photometry obtained with the 1-meter Swope telescope at Las Campanas
Observatory (LCO) using essentially the same instrument\footnote{Between observations of
SN~2013a and SN~2017cbv,
the direct camera CCD was upgraded from the original SITe3 to e2V. This introduces a
change in the zero-points of the filters, but leaves their relative shapes 
nearly identical.} and filters. SN~2013aa was also observed
in the optical by \citet{Graham:2017} and SN~2017cbv was observed in the optical by
\citet{Sand:2018}, both using the Las Cumbres Global Telescope (LCOGT) facilities.
High-quality, NIR photometry is available for both objects, though from different
telescopes. Both SNe~Ia appear to be normal with respect to decline-rate and 
have colors consistent with minimal to no reddening. Spectra of the SNe taken
at similar epochs indicate they are not only siblings, but also spectroscopically
very similar. SN~2017cbv is also unusual in having a very conspicuous "blue bump"
early in its light curve \citep{Hosseinzadeh17}, which has only been seen
in one other case \citep{Marion:2016}. Lastly, NGC~5643
is close enough to have its distance determined by primary methods, such as
Cepheid variables and the Tip of the Red Giant Branch (TGRB), making it an
important anchor for measuring the Hubble constant.

\section{Observations} \label{sec:observations}

In this section we briefly describe the observations of both SNe~Ia,
the data reduction methods, and the photometric systems involved.

\begin{figure}
    \centering
    \includegraphics[width=0.5\textwidth]{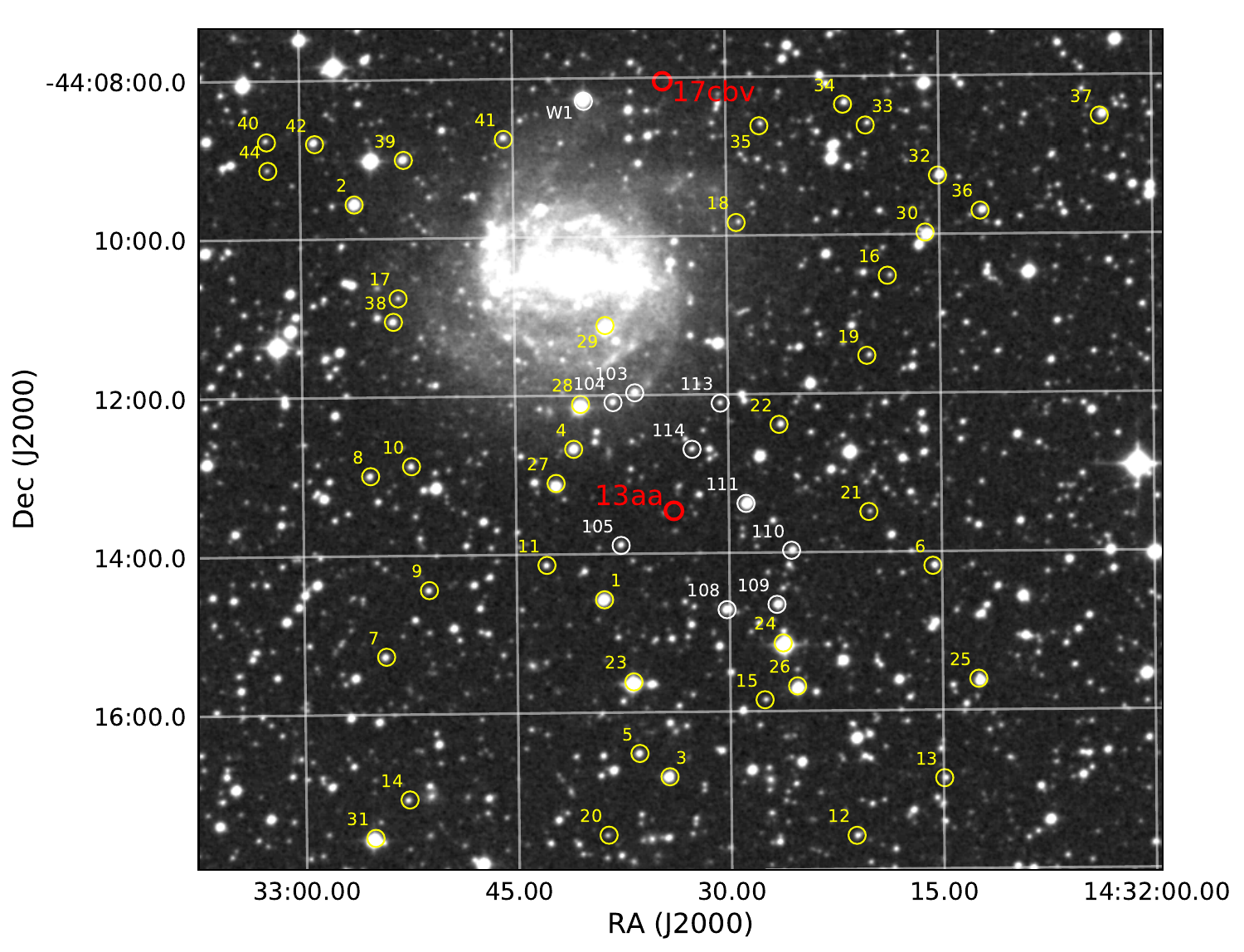}
    \caption{The field of SN~2013aa and SN~2017cbv. The image was taken on the du Pont
    telescope using the direct CCD camera through the $B$ filter. The locations
    of the two SNe are labeled
    in red. The local sequence stars are plotted as yellow circles and labeled with
    their identification numbers. North is up, East is to the left. 
    }
    \label{fig:field}
\end{figure}

\subsection{Photometry} \label{sec:photomery}
SN~2013aa was observed as part of the \textit{Carnegie Supernova Project II}
\citep[hereafter CSP-II]{Phillips:2018,Hsiao19}.
Optical imaging was obtained with the Swope
telescope equipped with the direct SITe3 CCD imager and a set of  $ugriBV$ filters. NIR photometry
was obtained using the 2.5-meter du Pont telescope with a set of $YJH$ filters.
The observing procedures, data reduction, and photometric systems are
outlined in other CSP papers \citep{Krisciunas:2017,Phillips:2018}. 

SN~2017cbv was observed as part of the Swope Supernova Survey 
\citep[][Rojas-Bravo et~al., in prep]{Coulter:2017}.
Similar to SN~2013aa, optical photometry was also 
obtained on the Swope telescope with the direct camera, but with an upgraded e2V CCD,
which has been fully characterized by the CSP-II \citep{Phillips:2018}.  While the 
surveys employed slightly different strategies for target selection, this does not 
affect the observations presented here.  The exact pointings for SN~2013aa and 
SN~2017cbv observations differ slightly to position each SN near the center of 
a chip.  However because of the large Swope field of view, there are typically 25 
local standard stars in common between the images, which we use to calibrate
all SN photometry.

We emphasize that the optical photometry of both SNe and the 
NIR photometry of SN~2013aa are given in the CSP natural system. Using a
long temporal baseline of observations at LCO, the CSP has determined
the color terms that transform the instrumental magnitudes obtained
on the Swope and du Pont telescopes into the standard magnitudes of \citet{Landolt:1992}
($BV$), \citet{Smith:2002} ($ugri$), \citet{Persson:1998} ($JH$) and
\citet{Krisciunas:2017} (Y).
Using these
color terms in reverse, we produce natural system magnitudes of our
standards and local sequence stars, which are then used to
differentially calibrate the photometry of each SN. This greatly 
simplifies the procedure of
transforming to another photometric system so long as the 
filter functions used to obtain science images are accurately measured. 

The NIR photometry of SN~2017cbv was obtained by
\citet{Wee:2018} using ANDICAM on the SMARTS 1.3-meter telescope
at the Cerro Tololo Inter-American Observatory (CTIO). They used
standard observing procedures, and tied their optical and NIR SN photometry to
a single local sequence star. Comparing
our $B$- and $V$-band photometry after computing S-corrections
\citep{Stritzinger2002}, we note a systematic difference of order
0.1 -- 0.2 mag, our photometry being dimmer. This appears to stem from 
the photometry of their
single local sequence star (star 1 in \citealt{Wee:2018}, star
W1 in Figure ~\ref{fig:field}), which is very
red ($B-V = 1.2$ mag) and comparatively bright ($V \sim 13$ mag). 
The red color leads to a relatively large color-term correction to
transform the instrumental magnitude to standard and may introduce a
large systematic error. In fact,
this star is bright enough to be saturated in all but our shallowest
exposures, so was not used to calibrate the Swope photometry. 
SN2017cbv was also observed independently with ANDICAM by a
separate group and we find their data to be in agreement with the CSP
data (Wang et al, in preparation).

The local sequence star used by \citet{Wee:2018} to calibrate the NIR 
photometry, in contrast, has colors more consistent with the 
\citet{Persson:1998}
standards ($J-H \sim 0.3$ mag). Furthermore, they have used the CSP-I
$Y$-band calibration from \cite{Krisciunas:2017}, with no color term 
applied, making their $Y$-band photometry nearly identical to the CSP
natural system, differing only in the shapes of the $Y$
bandpasses. Unfortunately, there is no overlap between our RetroCam
and the ANDICAM fields, making a direct comparison of local sequence
stars impossible.

\begin{figure*}
    \centering
    \includegraphics[width=0.35\textwidth]{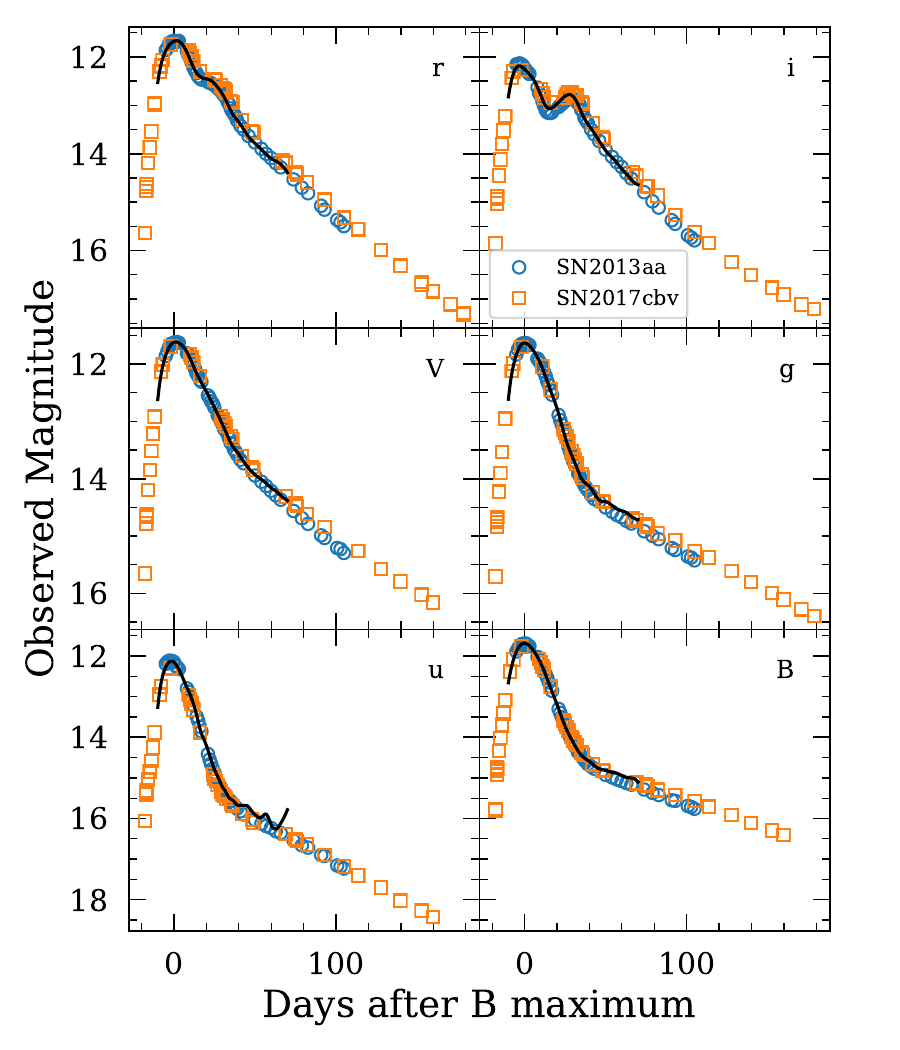}
    \includegraphics[width=0.3\textwidth]{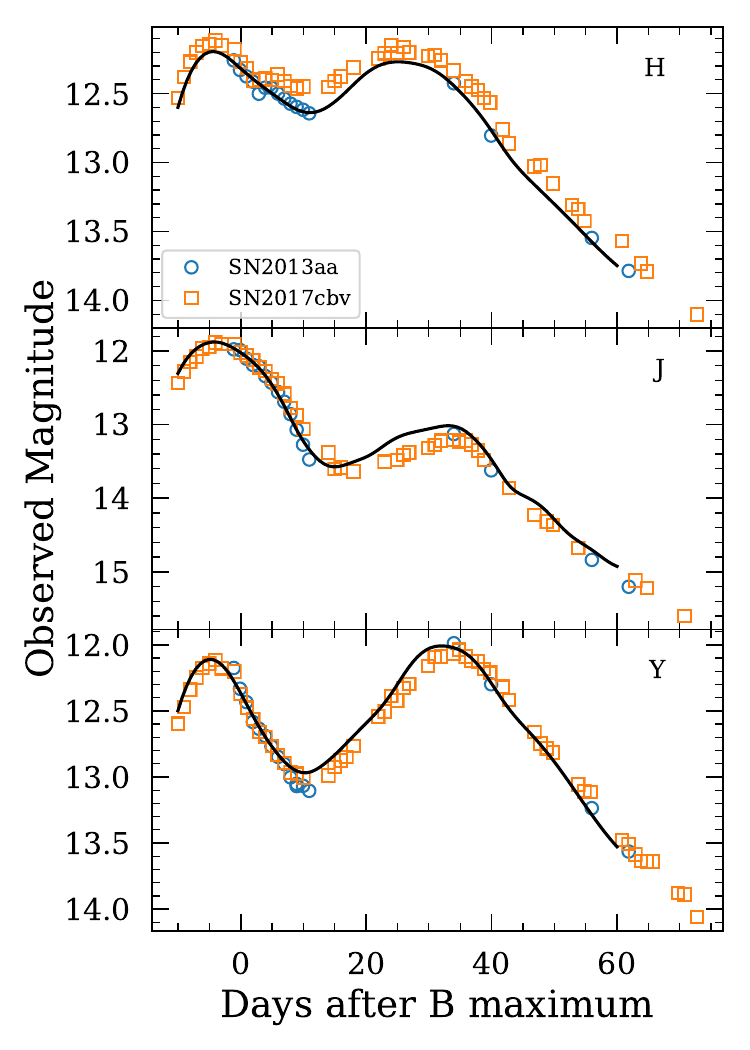}
    \includegraphics[width=0.3\textwidth]{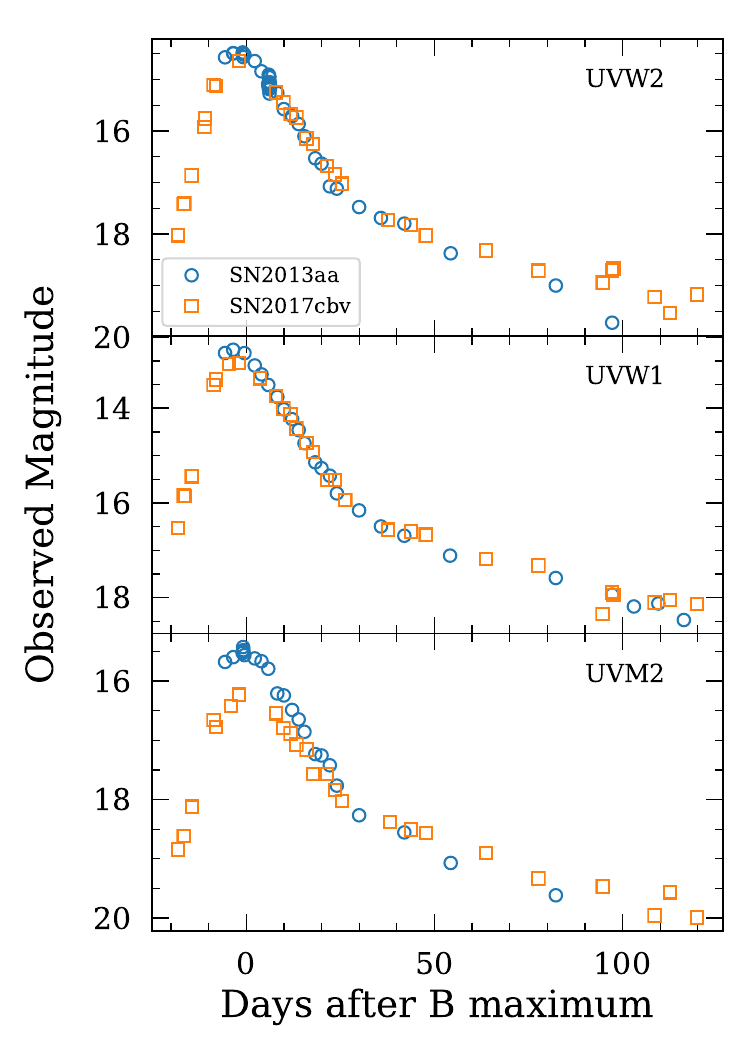}
    \caption{Optical, NIR and UV light curves of SN~2013aa and SN~2017cbv.
    The blue points correspond to SN~2013aa and the orange points
    correspond to SN~2017cbv. The NIR photometry of SN~2017cbv was obtained from images taken with ANDICAM. 
    The UV points are from SWIFT. The solid black curves are \texttt{SNooPy} fits to SN~2013aa. 
    }
    \label{fig:LCs}
\end{figure*}

Lastly, the \textit{Neil Gehrels Swift Observatory}  observed both SN~2013aa and SN~2017cbv as part of the
\textit{The	Swift Optical/Ultraviolet Supernova	Archive} (SOUSA; \citealt{Brown:2014}).
Details of the photometric data reduction and calibration are given in
\citet{Brown:2009}.

\startlongtable
\begin{deluxetable}{cccc}
\tablewidth{\textwidth}
\tablecaption{A log of the visual-wavelength and NIR spectra presented in this work. \label{table:NIR}}
\tablehead{
\colhead{SN}&
\colhead{$T_{spec}$\tablenotemark{a}}&
\colhead{Phase\tablenotemark{b}}&
\colhead{Instrument/Telescope}\\
\colhead{}&
\colhead{JD$-$2,400,000}&
\colhead{days}&
\colhead{}}
\startdata
  \multicolumn{4}{c}{\bf Optical}\\ 
   \hline
   SN\,2013aa & 56339.4 & $-$3.8 & MIKE/Magellan\\
   SN\,2017cbv & 57888.3 & +47.7 & MIKE/Magellan\\
  SN\,2017cbv&57853.5&$+$13.0&WiFeS/ANU 2.3-m\\
   SN\,2017cbv&57890.5&$+$50.0&WiFeS/ANU 2.3-m\\
    \hline
  \multicolumn{4}{c}{\bf NIR}\\ 
 \hline
    SN\,2013aa&56357.90&+14.7&FIRE/Baade\\
SN\,2017cbv&57857.64&+17.1&FIRE/Baade\\
\enddata
\tablenotetext{a}{Time of spectral observation.}
\tablenotetext{b}{ Phase of spectra in rest frame relative to the epoch of $B$-band maximum.}
\end{deluxetable}

\subsection{Spectroscopy}\label{sec:spectra}
Visual-wavelength spectra of SN~2017cbv were obtained from \citet{Hosseinzadeh17}, 
while those of SN\,2013aa come from WISeREP \citep{Yaron12}, with the exception of two
spectra that were obtained with WiFeS \citep{Dopita2007}.
WiFeS is an Integral Field Unit (IFU) on the 2.3-meter Australian National University
telescope located at the Siding Spring Observatory. The WiFeS IFU has a 
$25\times38$ arcsec field of view. Two gratings were used, for the blue and red cameras
respectively, with a resolution of $R=3000$ along with a dichroic at 5600~\AA. 
The data were reduced using pyWiFeS \citep{Childress2014}. 
The spectra were color-matched to the multi-band photometry to ensure accurate flux calibration.

We also present NIR spectra of both SN\,2013aa and SN\,2017cbv
obtained with the FIRE spectrograph on the 6.5-meter Magellan Baade telescope at LCO.
The spectra were reduced and corrected for telluric features following the procedures described 
by \citet{Hsiao19}.
Additionally, high-resolution spectra of SN~2013aa and SN~2017cbv were obtained  using the 
Magellan Inamori Kyocera Echelle \citep[MIKE;][]{Bernstein:2003}. The data were reduced
using the MIKE pipeline \citep{Kelson:2003} and procedures outlined in \citet{Simon:2010}.
A  journal of spectroscopic observations is  provided  in Table~\ref{table:NIR}. 
 
\section{Results}

We present here the analysis of the data from the previous section, including photometric
classification and the distances derived from 
common template light curve fitters. The spectra are used to delve into the physical
characteristics of the explosions and progenitors.

\begin{deluxetable}{lll}
\tablecaption{Comparison of LC parameters\label{tab:LCparams}}
\tablehead{
   \colhead{Parameter} & \colhead{SN~2013aa} & \colhead{SN~2017cbv} \\
   }
\startdata
   \multicolumn{3}{c}{Spline Fits}\\
   $t_{Bmax}$ (MJD)            & 56343.20(07) & 57840.54(15)\\
   $\Delta m_{15}(B)$ (mag)    & 0.95(01)      & 0.96(02) \\
   $s^D_{BV}$                    & 1.11(02)          & 1.11(03) \\
   $B_{max}$  & 11.094(003)       & 11.118(011)\\
   $V_{max}$                   & 11.143(004)       & 11.173(010)\\
   $B_{max}-V_{max}$           & -0.048(005)       & -0.056(015)\\
   \multicolumn{3}{c}{SNooPy\tablenotemark{a}} \\
   $t_{Bmax}$ (MJD)   & 56343.42(34) & 57840.39(34)\\
   $s_{BV}$           & 1.00(03)          & 1.12(03) \\
   $E(B-V)$ (mag)     & -0.03(06)  &  0.03(06) \\
   $A_V$ (mag)        & -0.06(12)  &  0.06(12) \\
   $\mu$ (mag)        & 30.47(08)  &  30.46(08) \\
   \multicolumn{3}{c}{MLCS2k2}\\
   $t_{Bmax}$ (MJD)   & 56343.40(11) days & 57839.79(06) \\
   $\Delta$ (mag) & -0.09(02) & -0.26(02) \\
   $A_V$ (mag) & -0.04(05) & 0.23(05) \\
   $\mu$ (mag)\tablenotemark{b} & 30.56(04) & 30.46(04) \\ 
   \multicolumn{3}{c}{SALT2}\\
   $t_{max}$ (MJD) & 56343.95(03) & 57840.66(03)\\
   $x_0$          & 0.71(0.01)    & 0.61(0.01) \\
   $x_1$          & 0.01(02)      & 1.30(17)\\
   $c$ (mag)      & -0.20(02)     & -0.02(03)\\
   $B_{max}$ (mag) & 11.01(02)    & 11.17(03)\\
   $\mu$ (mag)\tablenotemark{c}     & 30.62(04)    & 30.39(06)
\enddata
   \tablecomments{All magnitudes and colors are corrected for Milky-Way
   extinction $E(B-V) = 0.15$ mag based on \citet{Schlafly:2011}.}
   \tablenotetext{a}{Fit using the {\tt EBV2} model with $R_V=2$.}
   \tablenotetext{b}{Re-scaled from $H_0 = 65\ \mathrm{km\ s^{-1}\ Mpc^{-1}}$ to $H_0 = 72\ \mathrm{km\ s^{-1}\ Mpc^{-1}}$.}
   \tablenotetext{c}{Using ``JLA'' calibration of \citet{Betoule:2014} with $M_B^1$ re-scaled from
   $H_0 = 70\ \mathrm{km\ s^{-1}\ Mpc^{-1}}$ to $H_0 = 72\ \mathrm{km\ s^{-1}\ Mpc^{-1}}$}
\end{deluxetable}

\subsection{Decline Rate}\label{sec:phot_class}

Figure~\ref{fig:LCs} presents a comparison of the photometry of SN~2013aa and SN~2017cbv obtained with the
Swope, du Pont, SMARTS, and Swift telescopes. 
The light curves of the two objects are remarkably  similar, suggesting  their progenitors 
could also
be very similar. Table~\ref{tab:LCparams} lists pertinent 
photometric parameters estimated via the methods described below.

The most straightforward photometric diagnostic to compare is the light curve decline-rate 
parameter $\Delta m_{15}(B)$
\citep{Phillips:1993}.
Using the light curve analysis package \texttt{SNooPy}
\footnote{Analysis for this paper was done with \texttt{SNooPy} version 2.5.3, available at
\url{https://csp.obs.carnegiescience.edu/data/snpy} or \url{https://github.com/obscode/snpy}}
\citep{Burns:2011},  the photometry is interpolated at day 15 in the 
rest frame of the SNe after applying K-corrections 
\citep{Oke:1968} and time
dilation corrections. This results in nearly the same decline rate
$\Delta m_{15}(B) \simeq 0.95$ mag, indicating both objects decline slightly
slower than the typical $\Delta m_{15}(B)$ value of 1.1 mag 
\citep[Fig.~10]{Phillips:2018}.

\begin{figure}
    \centering
    \includegraphics[width=0.45\textwidth]{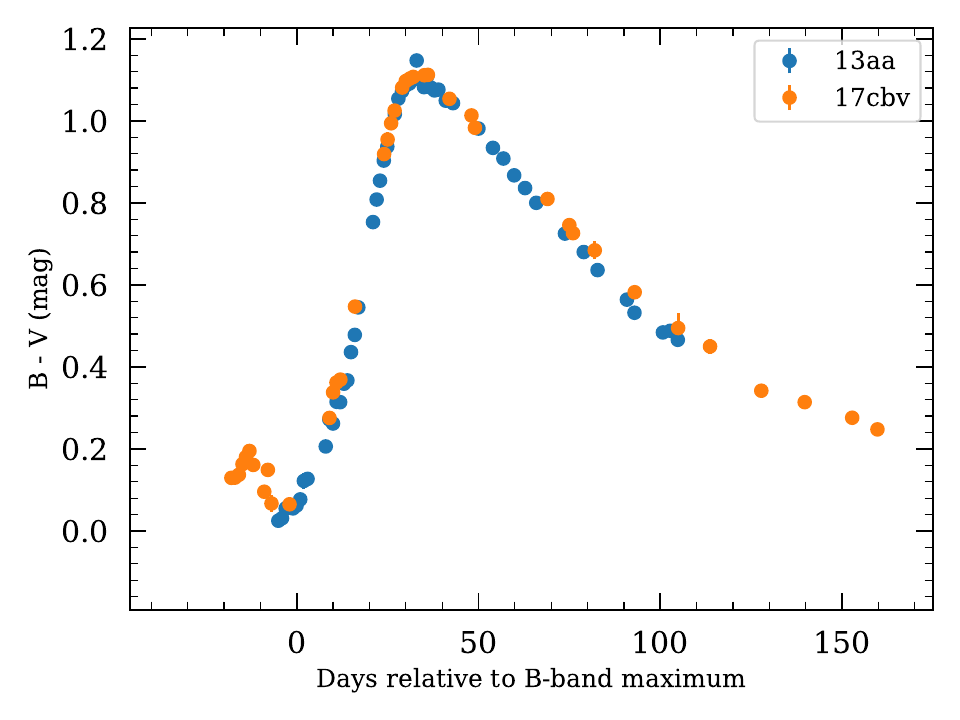}
    \caption{The $B-V$ color-curve evolution of SN~2013aa and SN~2017cbv. The time of $B-V$ maximum relative to $B$-band
    maximum divided by 30 days is defined as the color stretch,
    $s^D_{BV}$, which is nearly identical for these two SNe.}
    \label{fig:BV}
\end{figure}

Another way to characterize the decline-rate of the SNe directly
from photometry is to measure
the epoch that  the $B-V$ color-curves reaches their maximum point (i.e., when they
are reddest) in their evolution relative to the epoch of $B$-band maximum. Dividing
by 30 days gives the observed\footnote{
We shall use a super-script $D$ to distinguish between the directly-measured 
color-stretch $s^D_{BV}$ and the value inferred by multi-band template 
fits, $s_{BV}$.} color-stretch parameter $s^D_{BV}$. 
\citet{Burns:2014} showed
that the color-stretch parameter was a more robust way to classify SNe~Ia in terms of
light curve shape, intrinsic colors and distance \citep{Burns:2018}.
Figure \ref{fig:BV} shows the
$B-V$ color-curves for SN~2013aa and SN~2017cbv.
Fitting the $B-V$ color-curves with cubic splines,  we find identical
color-stretch values of 
$s^D_{BV} = 1.11 $ for both objects, again making them slightly slower
than typical SNe~Ia.

\begin{figure*}
    \centering
    \includegraphics[width=0.4\textwidth]{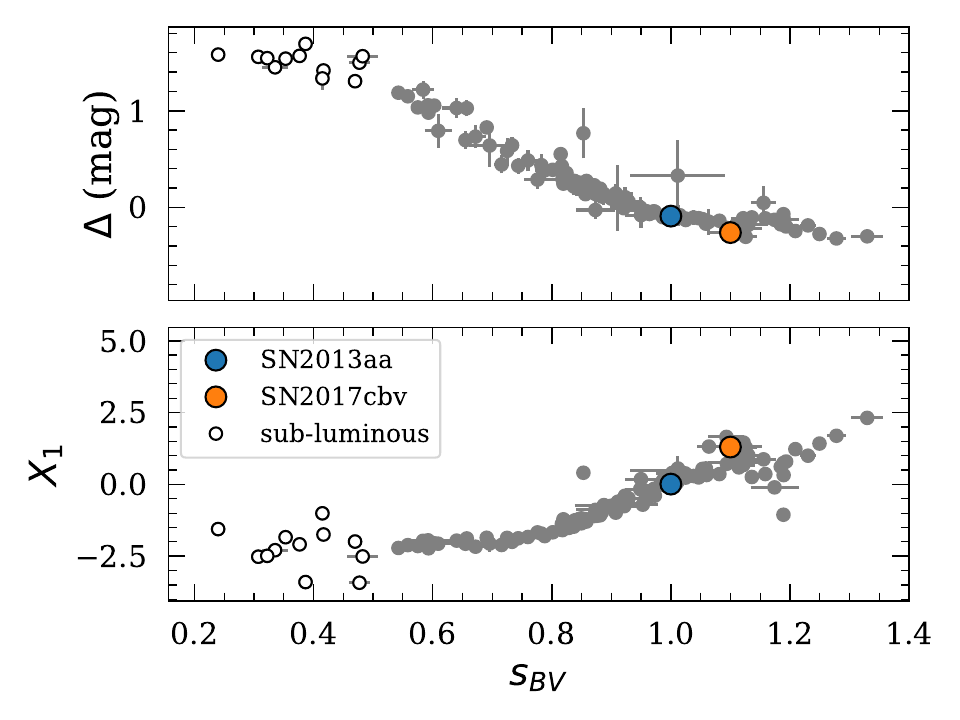}
    \includegraphics[width=0.4\textwidth]{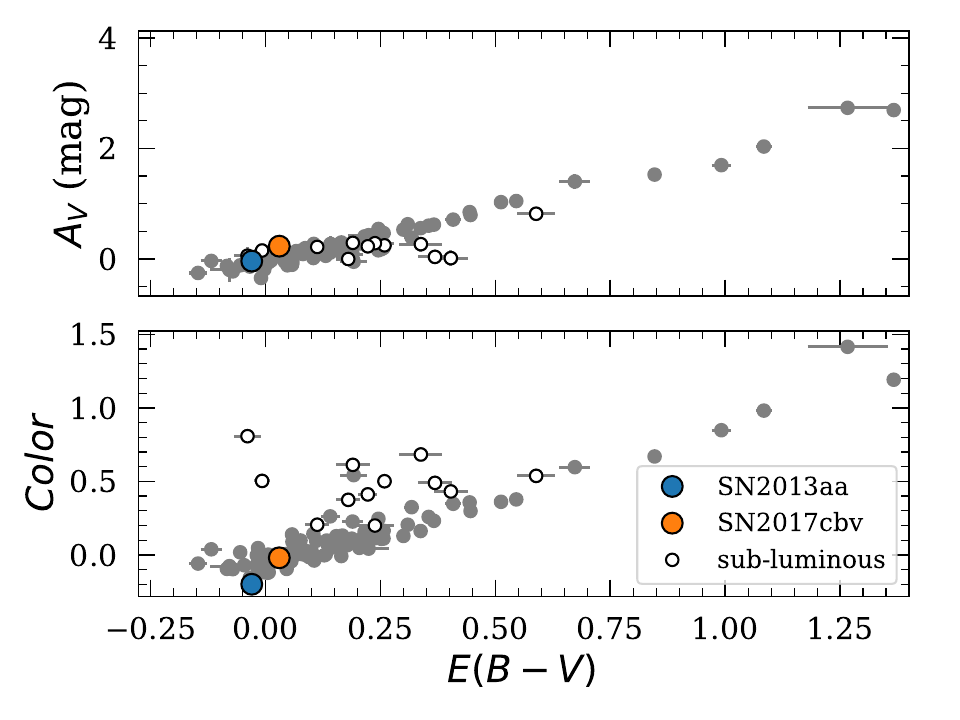}
    \caption{Comparison of parameters from three different SN~Ia
    light curve fitters for the CSP-I sample. (Left) Comparing
    the shape parameters, the \texttt{MLCS2k2}
    parameter $\Delta$ (top panel) and \texttt{SALT2} parameter
    $X_1$ (bottom panel) are plotted versus the \texttt{SNooPy} color-stretch
    parameter
    $s_{BV}$. (Right) Comparing color/reddening parameters,
    the \texttt{MLCS2k2} parameter $A_V$ (top panel) and
    \texttt{SALT2} color parameter (bottom panel) are plotted
    versus the \texttt{SNooPy} $E(B-V)$ parameter.
    SN~2013aa and SN~2017cbv are labeled with blue
    and orange circles, respectively. \label{fig:DeltaX1}}
\end{figure*}

For the purposes of determining distances for cosmology, it is
more common to fit multi-band photometry simultaneously, deriving
a joint estimate of the decline rate, color/extinction, and
distance. To this end,  the light curves of both SNe~Ia are fit
with 3 light curve fitting methods:  1) \texttt{SNooPy},
2) \texttt{SALT2}
\footnote{\texttt{SALT} version 2.4.2 is available from
\url{http://supernovae.in2p3.fr/salt/} with updated
CSP photometric system files available at 
\url{https://csp.obs.carnegiescience.edu/data/filters}}
\citep{Guy:2007}, and 3) \texttt{MLCS2k2}
\footnote{\texttt{MLCS2k2} version 0.07 is available from 
\url{https://www.physics.rutgers.edu/~saurabh/mlcs2k2/} and updated CSP
photometric system files are available at 
\url{https://csp.obs.carnegiescience.edu/data/filters}} \citep{Jha:2007}. 
The results 
of these fits are listed in Table~\ref{tab:LCparams}. In terms of
the decline rate,  a slightly different picture emerges: all 
three fitters classify SN~2013aa as a faster decliner than SN~2017cbv.
In Figure \ref{fig:DeltaX1} we show how the different decline-rate parameters 
(\texttt{SNooPy}'s $s_{BV}$, \texttt{SALT2}'s $x_1$, and
\texttt{MLCS2k2}'s $\Delta$)
relate to each other for the CSP-I sample
\citep{Krisciunas:2017}. It is clear that the three decline rate parameters
for the two SNe follow the general trend and are therefore measuring the same
subtle differences in light-curve shapes that tell us SN~2013aa
is a faster decliner, despite having nearly identical $\Delta m_{15}(B)$ and
$s^D_{BV}$. This is because $\Delta m_{15}(B)$ is determined from only
two points on the $B$ light curve (maximum and day 15) and $s^D_{BV}$ is
determined from one point on the $B$ light curve and one point on
the $B-V$ color curve, whereas template fitters like \texttt{SNooPy}, 
\texttt{MLCS2k2} and \texttt{SALT2} use the {\em shapes} of multi-band light curves
to determine the decline rates. Looking closely at the color-curves
of both SNe in Figure \ref{fig:BV}, while the peaks are nearly identical,
SN~2017cbv rises more quickly prior to maximum and declines more slowly
after maximum.

\subsection{Extinction}\label{sec:extinction}
Both SN~2013aa and SN~2017cbv are located in the outer parts of
NGC~5643 on opposite sides, having a host offset of 17 kpc and 14 kpc,
respectively. We therefore expect minimal to no host-galaxy dust reddening for either object.
 However, NGC~5643  is
at a relatively low galactic latitude ($l = 15.03$\degr) with
a predicted Milky-Way color excess of $E(B-V)_{MW} = 0.15$ mag 
\citep{Schlafly:2011}.

\begin{figure}
    \centering
    \includegraphics[width=0.4\textwidth]{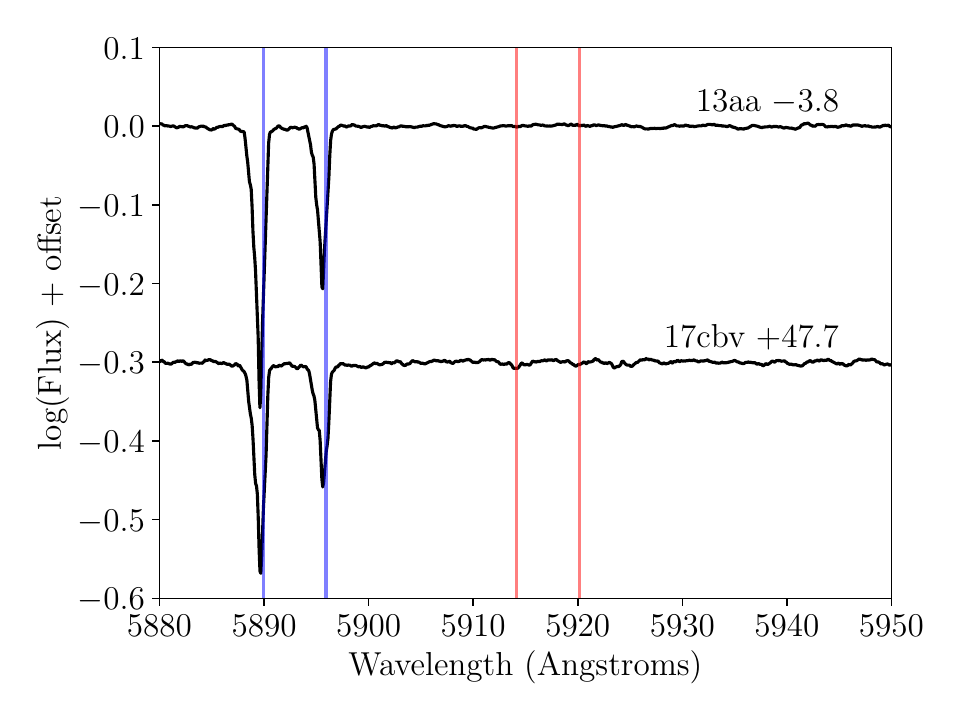}
    \caption{Continuum-normalized MIKE spectra of SN~2013aa and SN~2017cbv in the
    vicinity of
    the \NaI doublet. Absorption from the Milky-Way is clearly visible. The
    blue vertical lines denote the rest wavelengths of the \ion{Na}{1}~D2 $5890$~\AA\
    and \ion{Na}{1}~D1 $5896$~\AA\ lines, and the red lines show their location red-shifted
    to the systemic velocity of NGC~5643, i.e., 1200~kms. The names and 
    phases of the 
    spectra relative to the epoch of peak brightness are indicated to the right of each spectrum.
    \label{fig:NaI}}
\end{figure}

Figure \ref{fig:NaI} shows the continuum-normalized spectra of 
SN~2013aa and SN~2017cbv taken with MIKE
in the \ion{Na}{1}~D region. Absorption from the Milky-Way is clearly visible
in both cases,
while absorption at the systemic velocity of NGC~5643 is only detected
in the spectrum of SN~2017cbv. Measuring
the equivalent width of the combined \ion{Na}{1}~D lines and using the conversion
from \citet{Poznanski:2012}, we obtain an average Milky-Way reddening of
$E(B-V)_{MW} = 0.23\pm 0.16$ mag, somewhat higher than the \citet{Schlafly:2011}
value, but within the uncertainty. The equivalent width of the host \ion{Na}{1}~D
corresponds to $E(B-V)_{host} = 0.015 \pm 0.010$ mag. 
While the strength of \ion{Na}{1}~D absorption
has been shown to be a poor predictor of the amount of extinction
in SN~Ia
hosts, the absence of \ion{Na}{1}~D nevertheless seems to be a reliable 
indicator of a lack of dust reddening \citep{Phillips:2013}. In the remainder of this
paper the \citet{Schlafly:2011} value is adopted for the Milky-Way reddening
along with a reddening law characterized by  $R_V=3.1$, in order to correct 
the SN photometry for the effects of Milky-Way dust. 

The photometric colors of both objects are very 
blue at maximum, 
with SN~2017cbv being only slightly bluer. 
However, when the correlation between intrinsic color and
decline rate \citep[e.g.][]{Burns:2014} is taken into account, one infers slightly more 
extinction in SN~2017cbv than SN~2013aa.
\texttt{SALT2}, which does not estimate
the extinction but rather a rest-frame color parameter, gives a
bluer color ($c = -0.20$) for SN~2013aa than for SN~2017cbv ($c = -0.02$). In fact,
as can be seen in Figure \ref{fig:DeltaX1}, SN~2013aa's color parameter
is lower than our entire CSP-I sample. \texttt{MLCS2k2} provides estimates
of the visual extinction, $A_V$, and gives
a significant host extinction for SN~2017cbv ($A_V = 0.23\pm 0.05)$ mag,
or $E(B-V) = 0.12 \pm 0.03$ mag. However, like \texttt{SNooPy}, \texttt{MLCS2K2}
K-corrects its template light curves to fit the CSP filters and our
$u$ band is significantly different from Johnson/Cousins $U$, resulting in rather
large corrections. Eliminating $u$ from the fit brings the 
extinction estimate down to $A_V = 0.12 \pm 0.06$ mag or $E(B-V) = 0.06 \pm 0.03$ mag,
consistent to within the errors with the \texttt{SNooPy} value.

Given the positions of the two SNe, their lack of \ion{Na}{1}~D absorption at
the velocity of the host, and that \texttt{SNooPy} predicts zero color excess,
we conclude that SN~2013aa and SN~2017cbv experience minimal to no significant host-galaxy dust
extinction, making them ideal objects to improve upon the zero-point calibration of SNe~Ia. 

\subsection{UV Diversity}\label{sec:UV_diversity}
An outstanding feature of Figure \ref{fig:LCs} is the difference at peak in the single
Swift band $UVM2$ while the two other filters, $UVW1$ and $UVW2$, are consistent. 
This is likely due to the fact that both $UVW1$ and $UVW2$ have significant red leaks and
nearly 50\% of the flux comes from the optical light\footnote{See for example,
Fig.~1 of \citet{Brown:2010}.}. $UVM2$ is therefore a better indicator for
diversity in the UV, sampling the wavelength region $2000-3000$~\AA.

\begin{figure}
    \centering
    \includegraphics[width=0.4\textwidth]{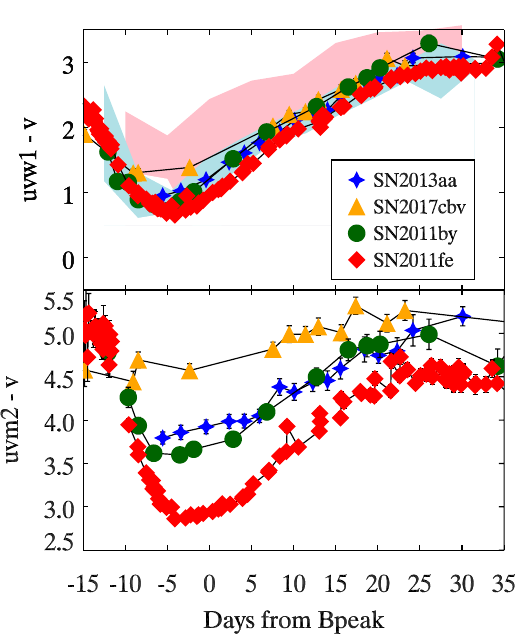}
    \caption{A comparison of the SWIFT UVOT $UVW1-V$ and $UVM2-v$ color curves.
    SN~2013aa, SN~2017cbv, and the twins SN~2011fe and SN~2011by are plotted
    with symbols. The red and blue shaded regions in the upper panel
    delimit the NUV-red and
    NUV-blue regions from \citet{Milne:2013}, respectively.}
    \label{fig:UVcomp}
\end{figure}

Figure~\ref{fig:UVcomp} shows the $UVW1-v$ and $UVM2-v$ colors as a function of time
for SN~2013aa and SN~2017cbv, as well as the two ``twins" SN~2011fe and SN~2011by. 
Two shaded regions are also shown indicating two populations of SNe~Ia: the NUV-blue 
and NUV-red as defined by \citet{Milne:2013}. They would seem to indicate that both
SN~2013aa and SN~2017cbv are NUV-blue. However, when comparing the $UVM2-v$ colors,
there is much more diversity and while SN~2013aa is similar to SN~2011by, SN~2017cbv
is much redder and SN~2011fe is much bluer, particularly near maximum light. This
underscores the issues with the red
leaks.
\citet{Milne:2013}
attribute these differences in NUV colors to differences in how close the burning front reaches
the surface of the ejecta. Alternatively, near-UV differences may be the result of varied
metal content in the ejecta (e.g, \citealp{Walker:2012,Brown:2019}), though no correlation 
has been found between the near-UV and the host galaxy metallicity \citep{Pan:2019,Brown_Crumpler:2019}.

\subsection{Spectroscopy}\label{sec:spec_class}
\subsubsection{Optical}

\begin{figure*}
    \centering
    \includegraphics[width=1.0\textwidth]{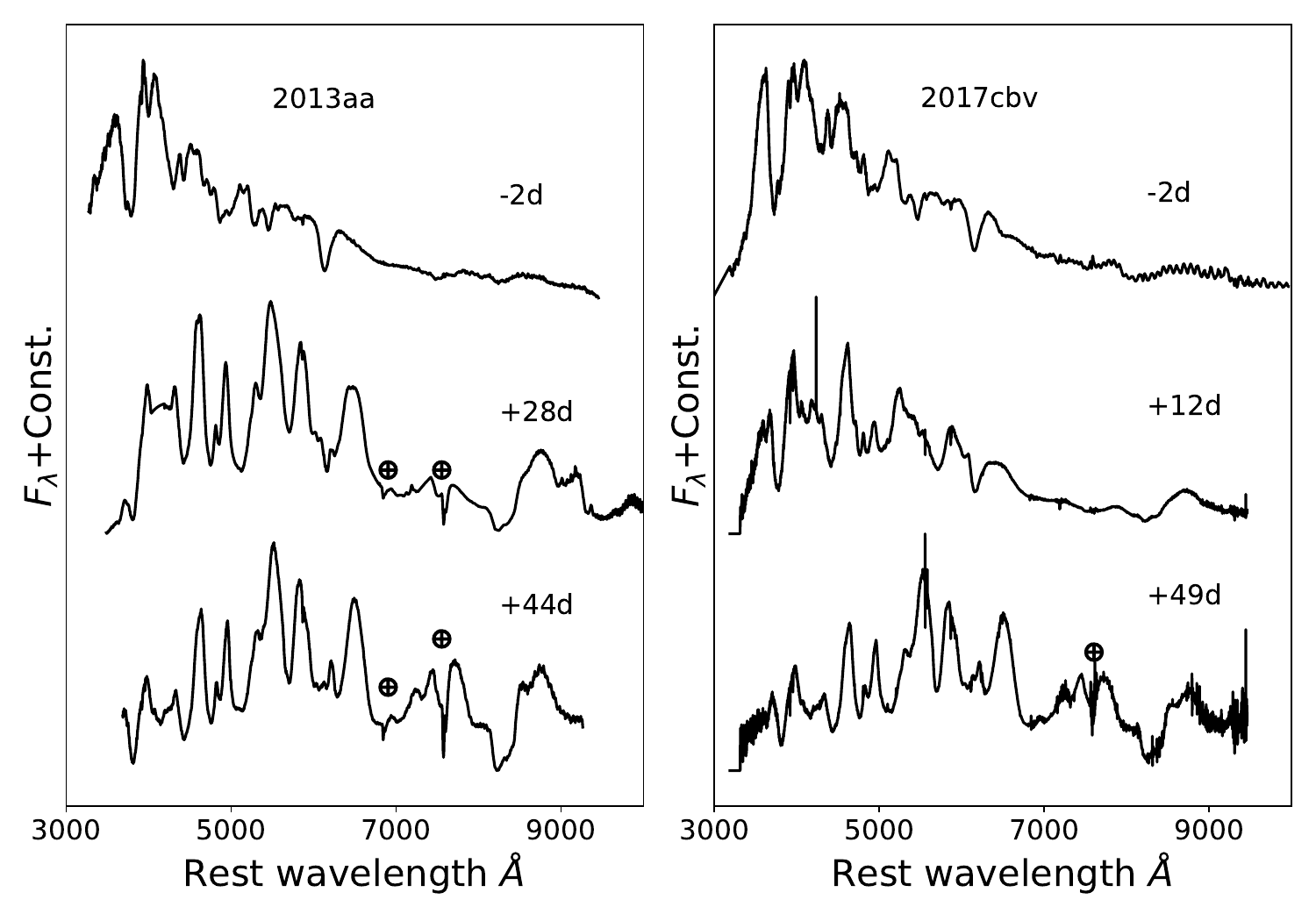}
    \caption{A rest frame time series of spectra for SN~2013aa (left) and SN~
    2017cbv (right). Phases are given relative to $B$-band maximum. Telluric 
    regions in the spectra are marked.}
    \label{fig:specall}
\end{figure*}

Figure~\ref{fig:specall} presents our  spectral time series of  SN~2013aa and
SN~2017cbv. Only  data obtained during the photospheric and transitional phase
are shown  as the nebular phase spectra will be the highlight of a future
study.  In the case of both objects,  three optical spectra were obtained
starting at $-2$~d with respect to $B$-band maximum. The last spectra of
SN~2013aa and SN~2017cbv were obtained on +44~d and +49~d, respectively. 

Doppler velocities and  pseudo-EW (pEW) measurements were calculated for each
object by fitting the corresponding features with a Gaussian function.  The
range over which the data were fit was manually selected, and  the continuum in
the selected region was estimated by a straight line. The velocity was then
measured by fitting the minimum of the Gaussian and the error was taken as the
formal error of the Gaussian fit. Finally,    pEW measurements were obtained
following the method discussed in \citet{Garavini07}. 

\begin{figure}
    \centering
    \includegraphics[width=0.4\textwidth]{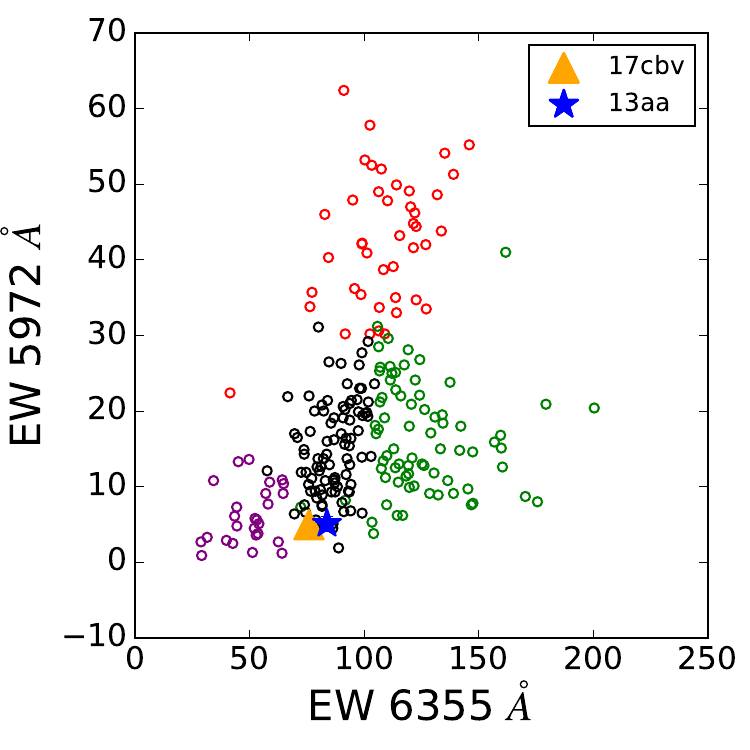}
    \caption{The Branch diagram for SNe Ia. SN\,2013aa (blue star) and SN\,
    2017cbv (orange triangle) are overlaid. The shallow silicon (purple), 
    core normal (black), broad line (green) and cool line (red) SNe are 
    plotted from  \citet{Blondin12}.}
    \label{fig:branch}
\end{figure}

The spectra of the SNe are characteristic of normal SNe~Ia.  
The $-2$\,d \ion{Si}{2} $\lambda$6355 Doppler velocity of SN~2013aa and SN~2017cbv are
$-$10,550$\pm$30\kms and $-$9,800$\pm$20\kms, respectively. This places them both as normal-velocity
SN~Ia in the \citet{Wang:2013} classification scheme. They are also both core normal (CN) in the 
\citet{Branch:2006} classification system, as demonstrated in the Branch diagram plotted in
Figure~\ref{fig:branch}. Although the two objects  are located close to the boundary between CN
and shallow silicon (SS) SNe~Ia.

\begin{figure*}
    \centering
    \includegraphics[width=0.95\textwidth]{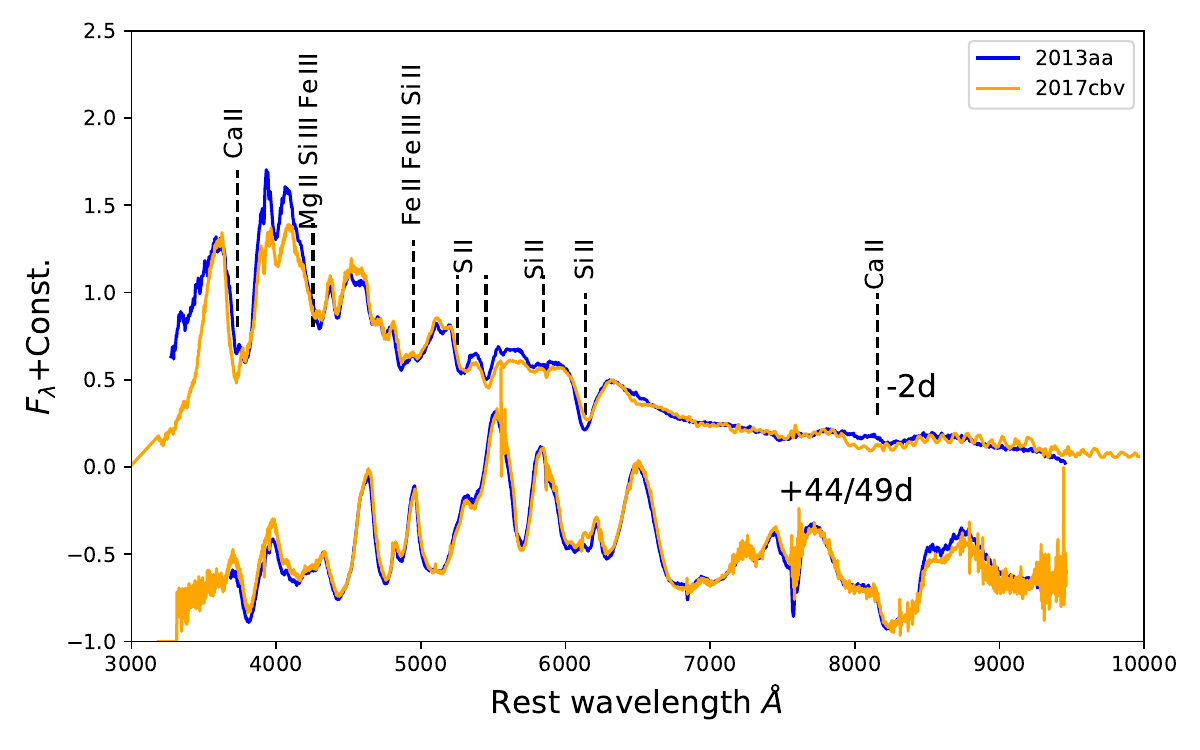}
    \caption{A comparison of the optical spectra of SN~2013aa and SN~2017cbv 
    at two epochs. Phases are given relative to $B$-band maximum.}
    \label{fig:spec}
\end{figure*}

Figure \ref{fig:spec} shows a comparison of the spectra of the two SNe obtained
at early ($-2$\,d) and later ($+44$\,d and $+49$\,d) phases.  At $-2$~d  the
spectra are nearly identical, exhibiting very similar line ratios and
ionization structures.  They contain  classic broad P~Cygni-like features typical of
SNe~Ia. They are also both dominated by doubly and singly ionized species, all
of which are labeled in Figure~\ref{fig:spec}.  The main difference between the
two objects at $-2$~d is the width of the \ion{Si}{2}  $\lambda$6355 feature,
where SN~2013aa (EW=84$\pm$1\AA) is broader than SN~2017cbv (EW=76$\pm$1\AA).
This suggests that SN~2013aa has a more extended Si region.  The spectra  of
SN~2013aa and SN~2017cbv  also look remarkably similar at +44~d and +49~d,
respectively.  At these epochs there is no longer a definitive photosphere and
emission lines  begin to emerge in the spectra. 

It has been hypothesized that using `twin' SN~Ia can improve their use as distance indicators \citep{Fakhouri:2015}.
 Given their photometric similarity, a natural question to address
is to what degree are SN~2013aa and SN~2017cbv  spectroscopic twins? Following the definition
of single phase twinness by \citet{Fakhouri:2015}, we compute their $\xi(p_i)$ parameter using
the $-2$~d spectra of both objects. $\xi$ is essentially a reduced $\chi^2$ statistic and 
therefore requires a good noise model that we  estimate by boxcar-averaging each spectrum
with a box size of 11 wavelength bins and subtracting these smoothed spectra from the 
originals. We also color-match each spectrum to match the observed photometry and only 
consider the wavelength range (4000--9500 \AA), corresponding to the wavelength coverage
of our $BVri$ photometry. The resulting value of $\xi(-2) = 1.7$ corresponds to 70\%
in the cumulative distribution from \citet{Fakhouri:2015}. In other words, 70\% of
SNe~Ia from their sample have a $\xi$ of 1.7 or less, indicating SN~2013aa and SN~2017cbv
are not spectroscopic twins by this metric. 
We note, however, that the two spectra were obtained
on different instruments. This  could introduce systematic errors that are not accounted for in
the noise model and could  lead to an over-estimate of $\xi$.

\subsubsection{NIR}

\begin{figure}
    \centering
    \includegraphics[width=0.45\textwidth]{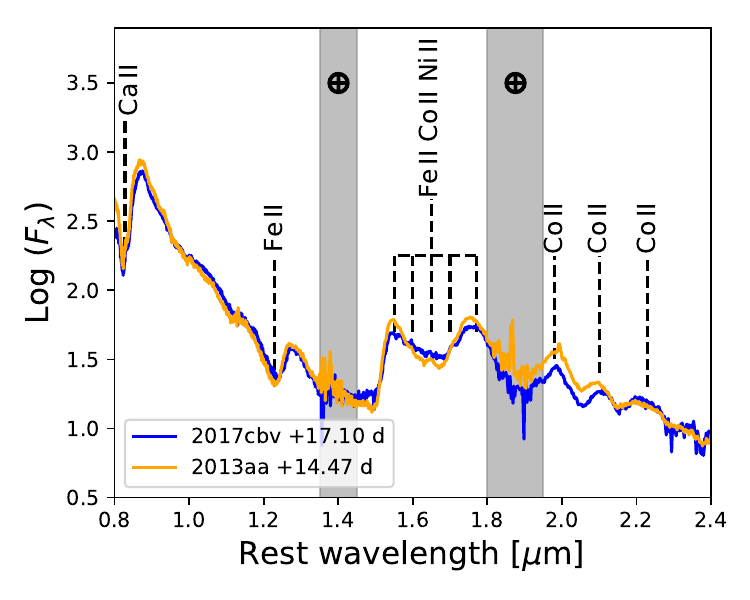}
    \caption{A spectral comparison between SN~2013aa and SN~2017cbv in the 
    NIR, at two different epochs. Phases are given in the legend.}
    \label{fig:NIRspec}
\end{figure}

NIR spectra of SN~2013aa and SN~2017cbv obtained at $+14.5$~d and +17.1~d,
respectively, are plotted in Figure~\ref{fig:NIRspec}.  The spectra are very
similar,  and are dominated by lines of iron group elements (see line IDs in
the figure).  A prevalent emission feature in the $H$-band region emerges in
all SNe~Ia by $+10$~d linked to  allowed emission lines of \CoII, \FeII\ and
\NiII\  produced from the radioactive decay of \Nifs\ which is located above
the photosphere \citep{Wheeler98,Hsiao:2015}.  \citet{Ashall19a} found a
correlation between the the outer blue-edge velocity, \ved, of this $H$-band
break region and \sBV. Furthermore, \citet{Ashall19b} found that \ved\ was a
direct measurement of the sharp transition between the incomplete Si-burning
region and the region of complete burning to \Nifs. \ved\  measures the \Nifs\
mass fraction between 0.03 to 0.10.

Using the method of  \citet{Ashall19a},  \ved\ is measured from the NIR spectra
of both objects.  At $+14.4$~d, SN~2013aa had a \ved\ of $-$13,600$\pm$300\kms,
and at $+17.1$~d SN~2017cbv had a \ved\ of $-$12,300$\pm$400\kms.  SN~2017cbv
has a lower value of \ved\ by 1300\kms, however this is likely to be due to the
fact that \ved\ decreases over time   \citep[see][]{Ashall19a}.

The fact that both SNe have similar values of \ved\ and absolute magnitude
indicates that they probably have a very similar total ejecta mass. As
explained in  \citet{Ashall19b}, for a given \Nifs\ mass, smaller ejecta
masses  produce larger values of \ved. However,  \ved\ is similar in both
SN~2017cbv and SN~2013aa, once the phase difference in the spectra is taken
into account.  Furthermore, the value of \ved\ obtained from SN~2017cbv is
consistent with similar SNe from \citet{Ashall19a}, as well as with
predictions of Chandrasekhar mass ($M_{Ch}$) delayed-detonation explosion
models  \citep{Hoeflich17,Ashall19b}. It should be noted that in
\citet{Ashall19a} and \citet{Ashall19b} the time period to measure \ved\ was
given as 10$\pm$3\,d.  However for normal-bright SNe~Ia the change in  \ved\
is slow, hence comparing to spectra at $+14.4$~d is still appropriate.

\section{Progenitors}

SNe~Ia are thought to originate from the thermonuclear disruption of
Carbon-Oxygen (C-O) white dwarfs (WD) in binary systems.  There are many
popular progenitor and explosions scenarios  \citep[see][for a recent
review]{Livio18}.  Two of the favored explosion scenarios, which can occur in
both the single and double degenerate progenitor system,  are the double
detonation (He-det) and delayed detonation scenarios (DDT). In the He-det
scenario a sub-$M_{Ch}$ WD accretes He from a companion, either a He star or
another WD with a He layer, the surface He layer detonates and drives a shock
wave into the WD producing a central detonation \citep{Livne95,Shen14,Shen18}.
In the DDT scenario, a WD accretes material until it approaches the $M_{Ch}$,
after which compressional heating near the WD center initiates a thermonuclear
runaway, with the burning first traveling as a subsonic deflagration wave and
then transitions into a supersonic detonation wave. 

It has been  predicted that SNe~Ia spectra can look similar at maximum light,
and their light curves can have the same shape, but their absolute magnitudes
can differ by \ab0.05\,mag \citep{Hoeflich17}. This should be the case for both
He-det (assuming the ejecta masses are similar) and DDT explosions.  This is
because at early times one of the dominant processes is how the \Nifs\ heats
the photosphere; that is, where the \Nifs\ is located with respect to the
photosphere.  However, once the photosphere is within the \Nifs\ region the
exact location of the \Nifs\  can differ and alter the light curve and spectra
after 30 days past maximum light \citep{Hoeflich17}.

Figure~\ref{fig:BV} shows that  the $B-V$ color curves for both objects are
very similar.  Furthermore, the optical spectra of SNe~2013aa and 2017cbv are
nearly identical at $\sim$+43~d  when the photosphere is well within the \Nifs\
region.  This demonstrates that these two objects are similar in the inner
regions, and have similar ignition mechanisms.  However, at $-2$~d the spectrum
of SN~2013aa has a broader  \ion{Si}{2} $\lambda$6355 feature indicating that
it has more effective burning and  a larger intermediate mass element (IME)
region. This implies that any small differences between the two objects is in
the outer layers. This could potentially be caused by differences in the main
sequence mass of the star before it produces the WD, which changes the C to O
ratio and the effectiveness of the burning in the outer layers
\citep{hk96,Hoeflich17,Shen18}.  This would alter the region between explosive
oxygen and incomplete Si burning in the ejecta, but not change the total \Nifs\
production and hence the luminosity, as is seen with the extended Si region in
SN~2013aa.  However, overall it is likely that the explosion mechanism for both
objects is very similar, and the observations of these SNe are largely
consistent with  the DDT and/or He-Det scenarios.

\section{Distance}\label{sec:distance}

\begin{deluxetable*}{llllllllll}
\tablecaption{SN~Ia siblings in the literature.\label{tab:siblings}}
   \tabletypesize{\footnotesize}
\tablehead{
   \colhead{Host} & \colhead{SN} & \colhead{$s_{BV}$} &  
   \colhead{$E(B-V)$} & \colhead{$\mu$} & \colhead{$\Delta \mu$} & 
   \colhead{Photometric system reference} \\
   \colhead{} & \colhead{} & \colhead{} & \colhead{mag} & \colhead{mag} &
   \colhead{mag} & \colhead{} 
   }
\startdata
NGC 105  & SN1997cw  & 1.30(04) & 0.29(07)    & 34.34(11) &          & CfA\tablenotemark{a}\\
         & SN2007A   & 1.01(04) & 0.24(06)    & 34.38(10) & 0.04(15) & CfA\tablenotemark{b}\\
         & SN2007A   & 1.10(02) & 0.24(06)    & 34.44(11) & 0.10(17) & CSP\tablenotemark{c}\\
NGC 1316 & SN2006mr  & 0.25(03) & 0.03(04)    & 31.26(04) &          & CSP\tablenotemark{c}  \citep{Burns:2018}\\
         & SN1980N   & 0.88(03) & 0.14(06)    & 31.27(09) & 0.01(10) & CTIO 1m\tablenotemark{d} - photographic\\
         & SN2006dd  & 0.93(03) & 0.09(06)    & 31.29(09) & 0.03(10) & ANDICAM \citep{Stritzinger:2010a}\\
         & SN1981D   & 0.77(05) & 0.05(09)    & 31.32(10) & 0.06(11) & CTIO 1m\tablenotemark{d}\\
NGC 1404 & SN2011iv  & 0.64(03) & $-$0.05(06) & 31.18(09) &          & CSP \\
         & SN2007on  & 0.58(03) & $-$0.06(06) & 31.59(10) & 0.41(13) & CSP \citep{Gall18}.\\
NGC 1954 & SN2013ex  & 0.92(03) & $-$0.01(06) & 33.64(09) &          & Swift UVOT\tablenotemark{e}\\
         & SN2010ko  & 0.57(04) & $-$0.07(07) & 33.92(14) & 0.28(17) & Swift UVOT\tablenotemark{e}\\
NGC 3190 & SN2002cv  & 0.85(04) & 5.40(09)    & 31.91(61) &          & Standard \citep{Elias-Rosa:2008}\\
         & SN2002bo  & 0.89(03) & 0.40(06)    & 32.03(13) & 0.12(62) & CfA\tablenotemark{f}\\
         & SN2002bo  & 0.94(03) & 0.43(06)    & 32.11(13) & 0.20(62) & KAIT\tablenotemark{g}\\
         & SN2002bo  & 0.92(03) & 0.42(06)    & 32.11(14) & 0.20(63) & Standard + LCO NIR \citep{Krisciunas:2004}\\
NGC 3905 & SN2009ds  & 1.05(03) & 0.07(06)    & 34.69(09) &          & CfA\tablenotemark{b}, PAIRITEL\tablenotemark{h}\\
         & SN2001E   & 1.02(04) & 0.47(06)    & 34.85(14) &  0.16(17)& KAIT\tablenotemark{g}\\
NGC 4493 & SN2004br  & 1.12(04) & 0.01(06)    & 34.82(08) &          & KAIT\tablenotemark{g}\\
         & SN1994M   & 0.88(04) & 0.17(06)    & 35.07(09) & 0.25(12) & CfA \citep{Riess:1999}\\
NGC 4708 & SN2016cvn & 1.25(12) & 0.91(13)    & 33.71(24) &          & Foundation \citep{Foley:2018a}\\
         & SN2005bo  & 0.79(03) & 0.28(06)    & 33.95(11) & 0.24(26) & CSP\tablenotemark{c}\\
         & SN2005bo  & 0.86(03) & 0.37(06)    & 33.96(12) & 0.25(27) & KAIT\tablenotemark{g}\\
NGC 5468 & SN1999cp  & 0.98(03) & 0.06(06)    & 33.10(08) &          & KAIT\tablenotemark{g}, 2MASS,  \citep{Krisciunas:2000}\\
         & SN2002cr  & 0.91(03) & 0.11(06)    & 33.17(08) & 0.07(11) & KAIT\tablenotemark{g}\\
         & SN2002cr  & 0.93(03) & 0.10(06)    & 33.21(09) & 0.11(12) & CfA\tablenotemark{f}\\
NGC 5490 & SN2015bo  & 0.41(08) & 0.11(13)    & 34.18(19) &          & Swift UVOT\tablenotemark{e}\\
         & SN1997cn  & 0.62(04) & 0.12(06)    & 34.57(10) & 0.39(21) & CfA\tablenotemark{a}\\
NGC 5643 & SN2017cbv & 1.09(04) & 0.08(06)    & 30.38(09) &          & Swift UVOT\tablenotemark{e}\\
         & SN2013aa  & 0.95(03) & $-$0.03(06) & 30.40(08) & 0.02(12) & LCOGT \citep{Graham:2017}\\
         & SN2017cbv & 1.13(03) & 0.03(06)    & 30.46(08) & 0.08(12) & CSP (This work)\\
         & SN2013aa  & 1.00(03) & $-$0.03(06) & 30.47(08) & 0.09(12) & CSP (This work)\\
         & SN2013aa  & 1.05(03) & $-$0.01(06) & 30.57(08) & 0.19(12) & Swift UVOT\tablenotemark{e}\\
NGC 6240 & PS1-14xw  & 0.95(04) & 0.26(06)    & 34.79(12) &          & Swift UVOT\tablenotemark{e}\\
         & SN2010gp  & 1.06(07) & 0.00(07)    & 35.22(12) & 0.43(17) & Swift UVOT\tablenotemark{e}\\
NGC 6261 & SN2008dt  & 0.87(05) & 0.49(09)    & 35.85(18) &          & CfA\tablenotemark{b}\\
         & SN2008dt  & 0.81(05) & 0.14(07)    & 35.87(15) & 0.02(23) & KAIT\tablenotemark{g}\\
         & SN2007hu  & 0.80(05) & 0.39(08)    & 35.91(14) & 0.06(22) & CfA\tablenotemark{b}\\
UGC 3218 & SN2011M   & 0.93(02) & 0.08(06)    & 34.37(08) &          & KAIT\tablenotemark{g}\\
         & SN2011M   & 0.85(05) & $-$0.01(07) & 34.37(10) & 0.00(14) & Swift UVOT\tablenotemark{e}\\
         & SN2006le  & 1.08(04) & $-$0.03(06) & 34.44(08) & 0.07(13) & KAIT\tablenotemark{g}\\
         & SN2006le  & 1.20(03) & $-$0.11(06) & 34.61(10) & 0.24(13) & CfA\tablenotemark{f}, PAIRITEL\tablenotemark{h}\\
UGC 7228 & SN2007sw  & 1.19(04) & 0.14(07)    & 35.25(10) &          & CfA\tablenotemark{b}\\
         & SN2012bh  & 1.11(04) & 0.10(06)    & 35.39(09) & 0.14(13) & PANSTARRS \citep{Jones:2018}\\
\enddata
\tablenotetext{a}{\citet{Jha:2006}.}
\tablenotetext{b}{\citet{Hicken:2012}.}
\tablenotetext{c}{\citet{Contreras:2010}.}
\tablenotetext{d}{\citet{Hamuy:1991}.}
\tablenotetext{e}{\citet{Brown:2014}.}
\tablenotetext{f}{\citet{Hicken:2009}.}
\tablenotetext{g}{\citet{Silverman:2012}.}
\tablenotetext{h}{\citet{Friedman:2015}.}
\end{deluxetable*}

With a heliocentric redshift $z_{hel} = 0.003999$ \citep{Koribalski:2004},
NGC~5643 is close enough to
have its distance determined independently, adding to the growing number of
hosts that can be used to calibrate the SN~Ia distance ladder for the purpose
of 
determining the Hubble constant.
And being siblings, SN~2013aa and SN~2017cbv
provide a consistency check on the reliability of SN~Ia distances in general.
Using the \citet{Burns:2014} calibration of
the Phillips relation and the light-curve parameters from SNooPy, 
we find distances of $\mu = 30.47\pm0.08$ mag and
$\mu = 30.46\pm 0.08$ mag for SN~2013aa and SN~2017cbv, respectively. 
The difference in distance modulus is $0.01 \pm 0.11$, and therefore is insignificant.
This compares well with the distance determined by \citet{Sand:2018}, who used \texttt{MLCS2k2}
on their own photometry to derive a distance of $\mu=30.45 \pm 0.09$.
\texttt{MLCS2k2} infers a distance of $\mu = 30.56 \pm 0.08$ mag for SN~2013aa and
$\mu = 30.46 \pm 0.08$ for SN~2017cbv, which is a difference of $0.10 \pm 0.11$ mag, so within the
uncertainty of the fitter. \texttt{SALT2}, however, infers distances of
$\mu = 30.62 \pm 0.11$ for SN~2013aa and $\mu = 30.39 \pm 0.12$ for SN~2017cbv, a
difference of $0.23 \pm 0.16$ mag. The larger differences for \texttt{SALT2} and
\texttt{MLCS2k2} are primarily due to differences in their color parameters
(\texttt{MLCS2k2} inferring SN~2017cbv to have significant $A_V$ and \texttt{SALT2}
inferring SN~2013aa to have very blue color).

Table \ref{tab:siblings} gives a list of the current sample of SNe~Ia siblings
in the literature. Using \texttt{SNooPy}, we have derived decline rates, extinctions
and distance estimates to each object. We then compare the inferred
host distances, which are tabulated in the column labeled $\Delta \mu$.
In all, it was found that 14 host
galaxies have hosted two SNe~Ia and one (NGC~1316) has hosted 4
SNe~Ia. The differences in distance estimate range from 0.02 mag to 0.43 mag.

Of particular interest are the siblings that were observed with the same
telescopes
and instruments, eliminating the systematic error of transforming photometry
from one system to another \citep{Stritzinger:2005}. In this regard, 
\textit{Swift} shows the greatest dispersion
among siblings with $\Delta \mu = 0.43$ mag for NGC~6240 and 
$\Delta \mu = 0.28$ mag for NGC~1954. In the
case of NGC~6240, the two SNe have color excesses that differ by 0.26 mag
and only the 
UVOT $B$ and $V$ filters could be reliably fit with \texttt{SNooPy}, requiring 
we assume the
typical $R_V=2$ for SNe~Ia \citep{Mandel:2009,Burns:2014}, rather than fit for it
using multi-band photometry. If $R_V$ is in fact higher by an amount
$\Delta R_V$ for this host, the discrepancy
would decrease by $\simeq 0.26\cdot\Delta R_V$.  As a result, if $R_V$ were as 
high as 4 in NGC~6240, the discrepancy would be eliminated.
In the case of NGC~1954, SN~2010ko
is a transitional Ia \citep{Hsiao:2013} which have been shown to be less reliable
as standard candles (see discussion of NGC~1404 below). This is also true of
NGC~5490, in which both SNe are transitional.

Further investigation comparing \textit{Swift} and CSP photometry has also shown
systematic errors between some SNe in common which can be as high as 0.15 mag.
Comparing the UVOT photometry of local sequence stars during the separate observing campaigns
of the sibling SNe does not show a significant difference (N. Crumpler, private 
communication).  The \textit{Swift} sibling SNe are also not located near bright stars or in bright
regions of the host which can invalidate the standard non-linearity corrections
\citep{Brown:2014}. Further investigation of these discrepancies is ongoing.

The Lick
Observatory Supernova Search (LOSS) observed two pairs of siblings in
NGC~5468 and UGC~3218 using the Katzman Automatic Imaging Telescope (KAIT). Both pairs
(SN~1999cp/SN~2002cr and SN~2006le/SN~2011M) exhibit similar decline rates and both pairs
had low reddening. The distance estimates for NGC~5468 differ by less than the error
($\Delta \mu = 0.07 \pm 0.11$), as is the case for UGC~3218 
($\Delta \mu = 0.07 \pm 0.13$). However, in other hosts where data were taken with
different telescopes, the KAIT distances tend to differ by 
$\Delta \mu \sim 0.2$ mag. LOSS publish their photometry in the standard
system \citep{Silverman:2012}, which is known to introduce systematic errors 
that are difficult
to correct and could be the cause of the larger discrepancy \citep{Stritzinger2002}.

The CfA supernova survey observed two pairs of siblings: one in NGC~6261 (SN~2008dt and
SN~2007bu) and
another in NGC~105 (SN~1997cw and SN~2007A). In both cases, the difference in the inferred
distance is well under
the uncertainties ($\Delta \mu = 0.06 \pm 0.22$ mag and $\Delta \mu = 0.04 \pm 0.15$ mag
respectively). In particular, both SN~2007A and SN~1997cw have moderately
high reddening $E(B-V)_{host} \sim 0.3$ mag, yet agree to within 2\%.
Again, however, when comparing siblings observed with different photometric systems,
the differences tend to be larger.

Lastly, prior to this paper, the CSP studied one pair of siblings in the host
NGC~1404, namely SN~2007on and SN~2011iv \citep{Gall18}. The difference in 
distance modulus obtained from the two objects was $\Delta \mu = 0.41$ mag, 
which is quite large for objects observed with the same facilities and 
hosted in the same galaxy. However, both objects have transitional
decline rates placing them between normal SNe~Ia ($s_{BV} > 0.6$) and 
fast-decliners ($s_{BV} < 0.4$), and it is likely that physical differences 
in their progenitors are responsible for their different peak luminosities \citep{Gall18}.

It is also worth re-examining NGC~1316 (Fornax A), which hosted 4 SNe~Ia. 
These were previously analyzed by \citet{Stritzinger:2010a}, who found that the three
normal SNe~Ia (SN~1980N, SN~1981D, and SN~2006dd) all agreed to within 0.03 mag,
but the extremely fast-declining SN~2006mr disagreed by nearly a magnitude. This
analysis was done using $\Delta m_{15}(B)$ as a light curve decline-rate parameter,
which was later shown to fail for fast-declining SN~Ia \citep{Burns:2014}. Using 
$s_{BV}$ instead leads to a SN~2006mr distance that is in complete agreement
with the other normal SNe~Ia \citep{Burns:2018}. This is the distance tabulated 
in Table~\ref{tab:siblings}. We therefore have siblings in NGC~1404 that seem to indicate
fast (or at least transitional) decliners are not as reliable, while NGC~1316
would indicate they are. If there is a diversity in progenitors at these decline
rates, then we may simply be seeing an increased dispersion in the Phillips
relation at the low $s_{BV}$ (high $\Delta m_{15}(B)$) end, or perhaps two different progenitor scenarios.
To know for sure will require an expanded sample of transitional SNe~Ia.

\begin{deluxetable}{llll}
   \tablecaption{Intrinsic dispersions in the sibling distances.
   \label{tab:sigma_SN}}
\tablehead{
   \colhead{Subsample} & \colhead{$\sigma_{SN}$} & \colhead{$N_{pair}$} & 
   \colhead{$N_{gal}$}
}
\startdata
   All pairs                       &  $0.14(02)$ & 34 & 15\\
   No Swift SNe                    &  $0.07(03)$ & 28 & 12\\
   $s_{BV} > 0.6$                  &  $0.12(02)$ & 25 & 12\\
   $s_{BV} > 0.6$ and No Swift SNe &  $<0.03$ (95\% conf.) & 21 & 11\\
\enddata
\end{deluxetable}

More quantitatively, we have 34 pairs of distances that can be compared,
including multiple observations of the same SN~Ia with different telescopes/
instruments.
Using a simple Bayesian hierarchical model, we can solve for an intrinsic 
dispersion $\sigma_{SN}$ in these distances, taking into account the 
photometric errors, errors in the SN~Ia calibration, and systematic errors
due to different photometric systems (see appendix B for details of this
modeling). Using all pairs, we derive $\sigma_{SN} = 0.14\pm 0.02$, however
this is dominated by the outliers mentioned above. If we eliminate the
Swift observations, the 
intrinsic dispersion reduces to $\sigma_{SN} = 0.07 \pm 0.03$,
or 3\% in distance. If we further remove the fast declining SNe
($s_{BV} < 0.6$), we obtain only an upper limit $\sigma_{SN} < 0.03$ at
95\% confidence. These
results are summarized in Table \ref{tab:sigma_SN}.

Within this landscape, we present two normal SNe~Ia siblings observed with the same telescope 
and nearly identical detector response functions. Unlike most of the siblings in Table \ref{tab:siblings},
SN~2013aa and SN~2017cbv have dense optical and NIR coverage, allowing  for
accurate measurement of the extinction, which is found to be consistent with minimal to no reddening
in both cases. The difference in distance modulus ($\Delta \mu = 0.01$) is less than
the uncertainties, as was the case with the CfA and KAIT siblings. Being at such
low redshift, we can expect the SN~Ia distances to differ from the Hubble 
distance modulus by about $\delta \mu = \frac{2.17 v_{pec}}{cz_{hel}} = 0.54$
mag for an assumed typical peculiar velocity of $v_{pec} = 300\ \mathrm{km\ s^{-1}}$.
With a CMB frame redshift $z_{cmb} = 0.0047$ and assumed Hubble
constant $H_0=72\ \mathrm{km\ s^{-1}\ Mpc^{-1}}$, the distance modulus is 
$\mu = 31.44$ mag, nearly a magnitude larger. Applying velocity corrections 
for Virgo, the Great Attractor, and the Shapley supercluster decreases the distance 
modulus to $\mu = 30.82$ mag. These differences in distance demonstrate the use
of standard candles such as SNe~Ia for determining departures from the 
Hubble expansion at low redshift. With a precision of 3\% in distances, we can
measure deviations from the Hubble flow at a level of $\delta v = 0.03 H_0 d$.
This corresponds to $\delta v \simeq 40 \mathrm{km\ s^{-1}}$ at the distance 
of the Virgo cluster ($20$ Mpc) and $\delta v \simeq 220 \mathrm{km\ s^{-1}}$ at
the distance of Coma ($100$ Mpc).

\section{Conclusions}\label{sec:conclusions}

The galaxy NGC~5643 is unique in that it has hosted two normal SNe~Ia that
have very similar properties and is close enough to have  its distance
determined independently. All photometric and spectroscopic diagnostics
indicate that SN~2013aa and SN~2017cbv are both normal SNe~Ia with minimal 
to no host-galaxy reddening.  This is also consistent with their positions in the
outskirts of the host galaxy. Both objects have been observed in the optical with the
same telescope and instruments and show a remarkable agreement in their
light curves.

Comparing the distances inferred by SN~2013aa and SN~2017cbv gives us the 
opportunity to test the relative precision of SNe~Ia as standard candles without
a number of systematics that typically plague such comparisons. Given that they occurred in the same
host galaxy, there is no additional uncertainty due to peculiar velocities. 
Having been observed by essentially the same telescope and instruments with the 
same filter set, there
is also minimal systematic uncertainty due to photometric zero-points or S-corrections. 
Finally, having no dust extinction, we eliminate the uncertainty due to
extinction corrections and variations in the reddening law 
\citep{Mandel:2009,Burns:2014}.
 When fitting multi-band photometry using \texttt{SNooPy}, \texttt{SALT2}, 
 and \texttt{MLCS2k2}, SN~2013aa is found to be characterized by a slightly faster
decline rate and bluer color at maximum.
The net result is a difference in distance that is insignificant compared to the
measurement errors, a similar situation as the other pairs of normal siblings
observed in multiple colors with the same instruments.

The similarity between the spectra and light curves of SN\,2013aa and SN\,2017cbv at all observed epochs 
suggest that they may have similar explosion mechanisms and  progenitor scenarios. 
However the differences between most leading explosion models are best 
seen at early and  late times. 
At these earliest epochs, SN\,2017cbv showed an early blue excess, but there were no data for SN\,2013aa.
Naturally the question is then: would SN\,2013aa also have shown this early 
blue excess? Although we cannot make this comparison for these
two objects,  
it is something that could be studied in the future using SN siblings. 
 
Some of the other questions that naturally arise for the future with 
twins and sibling studies are: if the spectral evolution were different,
e.g. one SN was high velocity gradient, and the other low velocity 
gradient, or 
if the SNe were different spectral sub-types at maximum light (e.g.  
shallow silicon vs. core normal vs. cool) would the distance calculated
for   each SN be different and, if so, does this point to different 
explosion 
mechanisms and progenitor scenarios? Also, with the advent of
Integral Field Spectrographs (IFS), we are beginning to study the local
host properties \citep{Galbany:2018}. Future IFS observations will allow us
to investigate any correlation between these local properties and differences
in inferred distance.

With the exception of the \textit{Swift} 
siblings, which had limited wavelength coverage,
we have four cases of normal SNe~Ia where most of the systematic errors are absent and
find the relative distance estimates to agree to within 3\%.  
Another host, NGC~1316, shows the same kind of consistency despite having multiple
photometric sources. This kind of internal precision rivals that of  Cepheid variables
\citep{Persson:2004}. The picture is
not as encouraging in the case of transitional- and fast-declining SNe~Ia, 
and more examples of these types of objects in the Hubble flow and/or 
additional pairs of such siblings  are required.
Comparing siblings from multiple telescopes also shows that we can expect 
disagreements on order of
5\% to 10\%, higher than typical systematic errors in the photometric systems themselves.
A likely reason for this is that these systematics in the photometric systems don't 
just affect the observed brightness, but also the observed \emph{colors}, which are
multiplied by the reddening slope $R_\lambda$ when correcting for dust.
It is therefore advantageous to use the reddest filters possible to avoid
this systematic when determining distances to SNe~Ia \citep{Freedman:2009,Mandel:2009,Avelino:2019}.

After submission of this paper, \cite{Scolnic:2020} released a preprint
detailing the analysis of sibling SNe~Ia from the DES survey. Unlike our
results, they find the dispersion among siblings to be no less than
the non-sibling SNe. Their sample, though, is quite different from ours. Their
SNe are photometrically classified while ours are spectroscopically classified.
They could therefore have peculiar SNe~Ia in their sample. Theirs is also a
higher redshift sample, ranging from $z = 0.228$ to $z = 0.648$. This results
not only in larger errors due to k-corrections, but also the fact that they
are measuring restframe wavelengths that are considerably bluer, on average,
than ours. 
The errors in the color corrections will therefore be larger. It is also
well-known that SNe~Ia show more diversity in the near-UV and UV
\citep{Foley:2008,Brown:2017}.

The analysis of sibling SNe gives us confidence that for appropriate cuts
in decline rate and when using well-understood photometric systems, relative
distances inferred from SNe~Ia are on par with primary indicators such as Cepheid
variables, but to greater distances. 

\acknowledgments
The work of the CSP-II has been generously supported by the National Science 
Foundation under grants AST-1008343, AST-1613426, AST-1613455, and AST-1613472. 
The CSP-II was also supported in part by the Danish Agency for Science and Technology
and Innovation through a Sapere Aude Level 2 grant. M. Stritzinger acknowledges
funding by a research grant (13261) from VILLUM FONDEN. T. D. is supported by an
appointment to the NASA Postdoctoral Program at the Goddard Space Flight Center,
administered by Universities Space Research Association under contract with NASA.
L.G. was funded by the European Union's Horizon 2020 research and innovation 
programme under the Marie Sk\l{}odowska-Curie grant agreement No. 839090.
This work has been partially supported by the Spanish grant 
PGC2018-095317-B-C21 within the European Funds for Regional Development (FEDER).
The UCSC team is supported in part by NASA grant NNG17PX03C, NASA/ESA Hubble 
Space Telescope program DD-14925 provided through STScI, NASA/Swift programs 
GI–1013136 and GI–1215205, NSF grants AST-1518052 and AST-1815935, the Gordon 
\& Betty Moore Foundation, the Heising-Simons Foundation, and by a fellowship 
from the David and Lucile Packard Foundation to R.J.F. 

\facilities{Swope (SITe3 and e2V imaging CCDs), Du Pont (Tek No. 5 imaging CCD, WFCCD),
Magellan:Clay (MIKE), Swift (UVOT),  La Silla-QUEST}

\software{\texttt{SNooPy} \citep{Burns:2014}, \texttt{SALT2} \citep{Guy:2007},
\texttt{MLCS2k2} \citep{Jha:2007}, \texttt{astropy} \citep{astropy:2018}, 
\texttt{matplotlib} \citep{Hunter:2007}}

\newpage
\clearpage
\appendix

\begin{deluxetable}{cccc}
   \tablecolumns{4}
   \tablecaption{Natural system photometry of SN~2013aa and SN~2017cbv. \label{tab:Phot}}
   \tablehead{
      \colhead{MJD} & \colhead{Filter} & \colhead{Magnitude} & 
      \colhead{Phase} \\
      \colhead{days} & & \colhead{mag} & \colhead{days}} 
   \startdata
\multicolumn{4}{c}{SN2013aa}\\
56338.360 & $B$ & $11.885$(006) & $-5.086$\\
56339.369 & $B$ & $11.811$(009) & $-4.078$\\
56340.384 & $B$ & $11.759$(007) & $-3.062$\\
56341.390 & $B$ & $11.707$(006) & $-2.056$\\
56342.386 & $B$ & $11.699$(006) & $-1.061$\\
56343.375 & $B$ & $11.692$(009) & $-0.071$\\
56344.392 & $B$ & $11.694$(009) & $0.946$\\
56345.381 & $B$ & $11.742$(015) & $1.935$\\
56346.394 & $B$ & $11.755$(013) & $2.948$\\
   \ldots & \ldots & \ldots & \ldots \\
\multicolumn{4}{c}{SN2017cbv}\\
57822.320 & $B$ & $15.805$(018) & $-18.340$\\
57822.322 & $B$ & $15.793$(016) & $-18.338$\\
57822.325 & $B$ & $15.771$(010) & $-18.335$\\
57823.175 & $B$ & $14.929$(016) & $-17.485$\\
57823.179 & $B$ & $14.912$(009) & $-17.481$\\
57823.286 & $B$ & $14.821$(015) & $-17.374$\\
57823.288 & $B$ & $14.829$(012) & $-17.372$\\
57823.388 & $B$ & $14.733$(015) & $-17.272$\\
57823.391 & $B$ & $14.768$(012) & $-17.269$\\
57824.381 & $B$ & $14.334$(007) & $-16.279$\\
   \ldots & \ldots & \ldots & \ldots \\
\enddata
\tablecomments{Table \ref{tab:Phot} is published in its entirety in the machine-readable format.
      A portion is shown here for guidance regarding its form and content.}
\end{deluxetable}

\startlongtable
\begin{deluxetable}{ccccccccchhhhhh}
   \tablecaption{Optical local sequence photometry for SN~2013aa and SN~2017cbv.
   \label{tab:LSPhot}}
   \tablehead{
      \colhead{ID} & \colhead{$\alpha$ (2000)} & \colhead{$\delta$ (2000)} & 
      \colhead{$B$} & \colhead{$V$} & \colhead{$u^\prime$} & \colhead{$g^\prime$} & 
      \colhead{$r^\prime$} & \colhead{$i^\prime$} & \nocolhead{$b$} & \nocolhead{$v$} & \nocolhead{$u$} &
      \nocolhead{$g$} & \nocolhead{$r$} & \nocolhead{$i$}}
   \startdata
   \multicolumn{15}{c}{SN~2013aa}\\
  1 & 218.161804 & -44.242950 & 14.452(003)& 13.842(003)& 15.363(007)& 14.114(003)& 13.710(003)& 13.553(003) & 14.415(003)& 13.859(003)& 15.306(007)& 14.120(003)& 13.713(003)& 13.553(003)\\
  2 & 218.234177 & -44.159451 & 15.595(004)& 14.391(003)& 17.675(018)& 14.960(003)& 13.991(003)& 13.608(003) & 15.522(004)& 14.436(003)& 17.550(018)& 14.974(003)& 13.997(003)& 13.609(003)\\
  3 & 218.143051 & -44.280270 & 15.458(004)& 14.664(004)& 16.676(011)& 15.037(004)& 14.440(003)& 14.244(003) & 15.410(004)& 14.688(004)& 16.601(011)& 15.045(004)& 14.443(003)& 14.244(003)\\
  4 & 218.170456 & -44.211262 & 15.951(004)& 15.146(004)& 17.178(014)& 15.520(004)& 14.910(003)& 14.702(004) & 15.902(004)& 15.172(004)& 17.102(014)& 15.529(004)& 14.913(003)& 14.702(004)\\
  5 & 218.151825 & -44.275311 & 16.909(007)& 16.231(005)& 17.785(019)& 16.551(005)& 16.051(006)& 15.836(006) & 16.868(007)& 16.254(005)& 17.728(019)& 16.558(005)& 16.054(006)& 15.836(006)\\
  6 & 218.065216 & -44.236328 & 17.081(007)& 16.307(005)& 18.054(024)& 16.668(005)& 16.075(005)& 15.833(005) & 17.034(007)& 16.335(005)& 17.990(024)& 16.676(005)& 16.079(005)& 15.833(005)\\
  7 & 218.226013 & -44.254551 & 17.351(009)& 16.450(006)& 18.776(043)& 16.890(005)& 16.165(006)& 15.915(006) & 17.296(009)& 16.481(006)& 18.689(043)& 16.900(005)& 16.169(006)& 15.915(006)\\
  8 & 218.230164 & -44.216572 & 17.646(011)& 16.794(007)& 18.770(047)& 17.211(006)& 16.505(006)& 16.218(005) & 17.594(011)& 16.827(007)& 18.698(047)& 17.221(006)& 16.510(006)& 16.219(005)\\
  9 & 218.213226 & -44.240639 & 17.705(011)& 16.924(007)& 18.904(051)& 17.330(007)& 16.686(006)& 16.359(006) & 17.657(011)& 16.957(007)& 18.832(051)& 17.339(007)& 16.691(006)& 16.360(006)\\
 10 & 218.218124 & -44.214561 & 17.924(013)& 16.953(007)& 19.079(094)& 17.427(007)& 16.625(006)& 16.233(005) & 17.865(013)& 16.995(007)& 19.003(094)& 17.438(007)& 16.631(006)& 16.234(005)\\
 11 & 218.178635 & -44.235592 & 18.010(014)& 17.062(008)& 19.413(111)& 17.541(007)& 16.749(007)& 16.460(006) & 17.952(014)& 17.097(008)& 19.327(111)& 17.552(007)& 16.754(007)& 16.461(006)\\
 12 & 218.088196 & -44.292938 & 17.881(013)& 17.003(008)& 19.238(069)& 17.413(007)& 16.724(007)& 16.439(007) & 17.827(013)& 17.036(008)& 19.154(069)& 17.423(007)& 16.729(007)& 16.440(007)\\
 13 & 218.062271 & -44.280991 & 18.048(015)& 17.130(009)& 19.354(086)& 17.551(008)& 16.614(006)& 16.520(007) & 17.992(015)& 17.165(009)& 19.271(086)& 17.564(008)& 16.615(006)& 16.520(007)\\
 14 & 218.219574 & -44.284592 & 18.302(018)& 17.345(010)& 18.988(182)& 17.796(009)& 16.938(008)& 16.597(007) & 18.244(018)& 17.388(010)& 18.933(182)& 17.808(009)& 16.943(008)& 16.598(007)\\
 15 & 218.114899 & -44.264271 & 18.517(021)& 17.703(013)& \ldots     & 18.103(011)& 17.409(010)& 17.141(010) & 18.467(021)& 17.736(013)& \ldots     & 18.113(011)& 17.413(010)& 17.142(010)\\
 16 & 218.077866 & -44.175289 & 17.639(011)& 17.619(011)& 17.998(024)& 17.557(008)& 17.715(012)& 17.879(016) & 17.638(011)& 17.604(011)& 17.978(024)& 17.555(008)& 17.712(012)& 17.879(016)\\
 17 & 218.221558 & -44.179260 & 18.492(021)& 17.833(014)& 19.464(122)& 18.191(013)& 17.676(012)& 17.444(012) & 18.452(021)& 17.856(014)& 19.405(122)& 18.198(013)& 17.680(012)& 17.444(012)\\
 18 & 218.122116 & -44.163929 & 18.675(026)& 17.909(017)& 19.428(179)& 18.332(017)& 17.731(014)& 17.434(013) & 18.628(026)& 17.936(017)& 19.378(179)& 18.340(017)& 17.736(014)& 17.435(013)\\
 19 & 218.084106 & -44.192081 & 18.747(025)& 17.768(013)& \ldots     & 18.260(013)& 17.385(010)& 17.053(009) & 18.687(025)& 17.809(013)& \ldots     & 18.272(013)& 17.390(010)& 17.054(009)\\
 20 & 218.161133 & -44.292419 & 18.954(031)& 18.137(019)& 19.034(118)& 18.496(017)& 17.919(015)& 17.663(015) & 18.904(031)& 18.164(019)& 19.009(118)& 18.504(017)& 17.923(015)& 17.664(015)\\
 21 & 218.083908 & -44.224899 & 18.568(022)& 17.833(014)& 19.159(068)& 18.190(012)& 17.583(011)& 17.315(010) & 18.523(022)& 17.863(014)& 19.114(068)& 18.199(012)& 17.587(011)& 17.316(010)\\
 22 & 218.110077 & -44.206421 & 18.025(014)& 16.957(007)& \ldots     & 17.484(007)& 16.515(006)& 16.143(005) & 17.960(014)& 17.004(007)& \ldots     & 17.498(007)& 16.521(006)& 16.144(005)\\
 23 & 218.153397 & -44.260262 & 12.893(003)& 12.342(003)& 13.746(006)& 12.558(003)& 12.191(003)& 12.092(003) & 12.859(003)& 12.356(003)& 13.691(006)& 12.563(003)& 12.193(003)& 12.092(003)\\
 24 & 218.109406 & -44.252369 & 12.242(002)& 11.677(003)& 13.116(005)& 11.900(003)& 11.510(003)& 11.388(003) & 12.208(002)& 11.694(003)& 13.060(005)& 11.905(003)& 11.512(003)& 11.388(003)\\
 25 & 218.052002 & -44.260151 & 15.225(004)& 14.612(005)& 16.126(010)& 14.893(004)& 14.456(004)& 14.290(004) & 15.188(004)& 14.631(005)& 16.069(010)& 14.899(004)& 14.459(004)& 14.290(004)\\
 26 & 218.105301 & -44.261421 & 14.268(003)& 13.596(003)& 15.183(007)& 13.898(003)& 13.421(003)& 13.235(003) & 14.227(003)& 13.617(003)& 15.124(007)& 13.905(003)& 13.424(003)& 13.235(003)\\
 27 & 218.175705 & -44.218399 & 15.070(003)& 14.348(003)& 16.076(009)& 14.679(003)& 14.154(003)& 13.958(003) & 15.026(003)& 14.371(003)& 16.012(009)& 14.686(003)& 14.157(003)& 13.958(003)\\
 28 & 218.168304 & -44.201851 & 14.148(003)& 13.100(003)& 15.629(008)& 13.591(003)& 12.759(003)& 12.414(002) & 14.084(003)& 13.140(003)& 15.535(008)& 13.603(003)& 12.765(003)& 12.415(002)\\
 29 & 218.160904 & -44.185322 & 13.640(003)& 12.977(003)& 14.600(006)& 13.272(003)& 12.783(003)& 12.610(002) & 13.600(003)& 12.998(003)& 14.539(006)& 13.279(003)& 12.786(003)& 12.610(002)\\
 30 & 218.066696 & -44.166279 & 14.534(003)& 13.849(003)& 15.485(008)& 14.157(003)& 13.672(003)& 13.488(003) & 14.492(003)& 13.870(003)& 15.424(008)& 14.164(003)& 13.675(003)& 13.488(003)\\
 31 & 218.229706 & -44.292591 & 13.372(003)& 12.630(003)& 14.468(009)& 12.950(003)& 12.394(004)& 12.216(003) & 13.327(003)& 12.654(003)& 14.398(009)& 12.958(003)& 12.397(004)& 12.216(003)\\
   \multicolumn{15}{c}{SN~2017cbv}\\
  1 & 218.161804 & -44.242950 & 14.403(014)& 13.812(006)& 15.315(023)& 14.091(007)& 13.650(008)& 13.518(010) & 14.349(014)& 13.830(006)& 15.278(023)& 14.093(007)& 13.650(008)& 13.515(010)\\
  2 & 218.234177 & -44.159451 & 15.602(030)& 14.372(006)& 17.465(086)& 14.944(006)& 13.963(007)& 13.601(011) & 15.490(030)& 14.420(006)& 17.389(086)& 14.949(006)& 13.963(007)& 13.593(011)\\
  3 & 218.143051 & -44.280270 & \ldots     & \ldots     & \ldots     & \ldots     & \ldots     & \ldots      & 15.370(019)& 14.690(015)& \ldots     & \ldots     & 14.661(021)& \ldots     \\
  4 & 218.170456 & -44.211262 & 15.955(037)& 15.132(006)& 16.968(055)& 15.499(007)& 14.888(006)& 14.673(008) & 15.880(037)& 15.160(006)& 16.924(055)& 15.502(007)& 14.888(006)& 14.668(008)\\
  5 & 218.151825 & -44.275311 & 16.934(023)& 16.223(007)& \ldots     & 16.509(009)& 16.027(011)& 15.814(011) & 16.869(023)& 16.248(007)& \ldots     & 16.511(009)& 16.027(011)& 15.809(011)\\
  6 & 218.065216 & -44.236328 & 17.087(057)& 16.291(007)& \ldots     & 16.652(006)& 16.044(006)& 15.814(008) & 17.015(057)& 16.321(007)& \ldots     & 16.655(006)& 16.044(006)& 15.809(008)\\
  7 & 218.226013 & -44.254551 & 17.370(050)& 16.429(007)& \ldots     & 16.868(007)& 16.132(007)& 15.905(009) & 17.284(050)& 16.462(007)& \ldots     & 16.872(007)& 16.132(007)& 15.900(009)\\
  8 & 218.230164 & -44.216572 & 17.667(115)& 16.784(008)& \ldots     & 17.203(008)& 16.467(006)& 16.209(006) & 17.587(115)& 16.820(008)& \ldots     & 17.207(008)& 16.467(006)& 16.203(006)\\
  9 & 218.213242 & -44.240639 & 17.794(031)& 16.917(009)& \ldots     & 17.325(008)& 16.618(007)& 16.356(007) & 17.714(031)& 16.952(009)& \ldots     & 17.329(008)& 16.618(007)& 16.350(007)\\
 10 & 218.218124 & -44.214561 & 17.925(066)& 16.944(009)& \ldots     & 17.427(008)& 16.553(007)& 16.231(006) & 17.836(066)& 16.988(009)& \ldots     & 17.431(008)& 16.553(007)& 16.224(006)\\
 11 & 218.178635 & -44.235592 & 18.077(045)& 17.057(010)& \ldots     & 17.538(009)& 16.731(007)& 16.458(007) & 17.984(045)& 17.094(010)& \ldots     & 17.542(009)& 16.731(007)& 16.452(007)\\
 12 & 218.088196 & -44.292938 & 17.890(013)& 17.000(010)& \ldots     & 17.393(009)& 16.693(008)& 16.428(008) & 17.809(013)& 17.036(010)& \ldots     & 17.397(009)& 16.693(008)& 16.422(008)\\
 13 & 218.062271 & -44.280991 & 18.020(070)& 17.121(011)& \ldots     & 17.538(010)& 16.792(008)& 16.515(008) & 17.938(070)& 17.159(011)& \ldots     & 17.542(010)& 16.792(008)& 16.509(008)\\
 14 & 218.219559 & -44.284592 & 18.347(109)& 17.329(012)& \ldots     & 17.789(012)& 16.906(009)& 16.595(009) & 18.254(109)& 17.374(012)& \ldots     & 17.793(012)& 16.906(009)& 16.588(009)\\
 15 & 218.114899 & -44.264271 & 18.552(162)& 17.694(015)& \ldots     & 18.088(014)& 17.378(011)& 17.134(011) & 18.474(162)& 17.729(015)& \ldots     & 18.092(014)& 17.378(011)& 17.129(011)\\
 16 & 218.077881 & -44.175289 & 17.642(079)& 17.599(013)& 17.307(135)& 17.549(009)& 17.705(013)& 17.862(017) & 17.638(079)& 17.583(013)& 17.314(135)& 17.548(009)& 17.705(013)& 17.865(017)\\
 17 & 218.221558 & -44.179260 & 18.511(065)& 17.811(016)& \ldots     & 18.191(015)& 17.666(013)& 17.453(013) & 18.447(065)& 17.833(016)& \ldots     & 18.194(015)& 17.666(013)& 17.448(013)\\
 18 & 218.122131 & -44.163929 & 18.738(066)& 17.904(019)& \ldots     & 18.312(020)& 17.707(016)& 17.428(015) & 18.662(066)& 17.934(019)& \ldots     & 18.315(020)& 17.707(016)& 17.422(015)\\
 19 & 218.084106 & -44.192081 & 18.811(081)& 17.771(016)& \ldots     & 18.242(015)& 17.373(010)& 17.056(009) & 18.716(081)& 17.815(016)& \ldots     & 18.246(015)& 17.373(010)& 17.049(009)\\
 20 & 218.161133 & -44.292419 & 18.918(010)& 18.115(022)& \ldots     & 18.479(020)& 17.886(015)& 17.660(017) & 18.845(010)& 18.143(022)& \ldots     & 18.482(020)& 17.886(015)& 17.655(017)\\
 21 & 218.083908 & -44.224899 & 18.616(052)& 17.807(016)& \ldots     & 18.186(014)& 17.563(012)& 17.312(011) & 18.542(052)& 17.838(016)& \ldots     & 18.189(014)& 17.563(012)& 17.307(011)\\
 22 & 218.110062 & -44.206421 & 18.074(040)& 16.951(009)& \ldots     & 17.485(009)& 16.486(006)& 16.132(006) & 17.972(040)& 17.002(009)& \ldots     & 17.490(009)& 16.486(006)& 16.124(006)\\
 23 & 218.153397 & -44.260262 & \ldots     & \ldots     & \ldots     & \ldots     & \ldots     & \ldots      & \ldots     & \ldots     & \ldots     & \ldots     & \ldots     & \ldots     \\
 24 & 218.109406 & -44.252369 & \ldots     & \ldots     & \ldots     & \ldots     & \ldots     & \ldots      & \ldots     & 11.627(019)& 13.046(022)& \ldots     & \ldots     & \ldots     \\
 25 & 218.052002 & -44.260151 & \ldots     & \ldots     & \ldots     & \ldots     & \ldots     & \ldots      & 15.160(011)& \ldots     & \ldots     & \ldots     & 14.615(021)& \ldots     \\
 26 & 218.105301 & -44.261421 & \ldots     & \ldots     & \ldots     & \ldots     & \ldots     & \ldots      & \ldots     & \ldots     & \ldots     & \ldots     & 14.088(021)& \ldots     \\
 27 & 218.175705 & -44.218399 & 15.052(039)& 14.325(006)& 16.031(029)& 14.641(006)& 14.109(007)& 13.930(008) & 14.986(039)& 14.349(006)& 15.989(029)& 14.644(006)& 14.109(007)& 13.926(008)\\
 28 & 218.168289 & -44.201851 & \ldots     & \ldots     & \ldots     & \ldots     & \ldots     & \ldots      & 14.036(016)& 13.116(006)& 15.422(024)& 13.566(015)& \ldots     & \ldots     \\
 29 & 218.160904 & -44.185322 & \ldots     & \ldots     & \ldots     & \ldots     & \ldots     & \ldots      & 13.577(013)& 12.985(007)& 14.550(020)& \ldots     & \ldots     & \ldots     \\
 30 & 218.066696 & -44.166279 & \ldots     & \ldots     & \ldots     & \ldots     & \ldots     & \ldots      & 14.447(018)& 13.852(006)& 15.422(024)& 14.125(007)& 13.630(008)& \ldots     \\
 31 & 218.229706 & -44.292591 & \ldots     & \ldots     & \ldots     & \ldots     & \ldots     & \ldots      & \ldots     & \ldots     & \ldots     & \ldots     & \ldots     & \ldots     \\
 32 & 218.062912 & -44.154339 & 15.588(032)& 14.885(005)& 16.648(043)& 15.191(006)& 14.674(006)& 14.474(007) & 15.524(032)& 14.910(005)& 16.604(043)& 15.194(006)& 14.674(006)& 14.470(007)\\
 33 & 218.084000 & -44.143681 & 18.297(058)& 17.534(014)& \ldots     & 17.930(012)& 17.319(011)& 17.081(010) & 18.228(058)& 17.562(014)& \ldots     & 17.933(012)& 17.319(011)& 17.076(010)\\
 34 & 218.090591 & -44.139259 & 17.739(050)& 16.692(008)& \ldots     & 17.168(008)& 16.353(007)& 16.056(006) & 17.644(050)& 16.731(008)& \ldots     & 17.172(008)& 16.353(007)& 16.050(006)\\
 35 & 218.115189 & -44.143639 & 18.473(040)& 17.633(015)& \ldots     & 18.026(014)& 17.341(011)& 17.100(011) & 18.397(040)& 17.666(015)& \ldots     & 18.029(014)& 17.341(011)& 17.095(011)\\
 36 & 218.050507 & -44.161770 & 16.711(047)& 15.877(006)& 17.595(113)& 16.258(007)& 15.607(006)& 15.319(007) & 16.635(047)& 15.912(006)& 17.555(113)& 16.261(007)& 15.607(006)& 15.313(007)\\
 37 & 218.015289 & -44.142010 & 16.065(065)& 15.369(006)& 17.025(057)& 15.694(007)& 15.156(005)& 14.981(007) & 16.002(065)& 15.393(006)& 16.985(057)& 15.697(007)& 15.156(005)& 14.977(007)\\
 38 & 218.223007 & -44.184189 & 16.707(068)& 16.111(006)& 17.566(111)& 16.368(007)& 15.957(007)& 15.811(008) & 16.653(068)& 16.130(006)& 17.530(111)& 16.370(007)& 15.957(007)& 15.808(008)\\
 39 & 218.219589 & -44.150181 & 16.727(034)& 15.770(006)& 17.759(123)& 16.229(009)& 15.456(006)& 15.135(007) & 16.640(034)& 15.809(006)& 17.713(123)& 16.233(009)& 15.456(006)& 15.128(007)\\
 40 & 218.259705 & -44.146191 & 18.403(139)& 17.674(014)& \ldots     & 18.035(012)& 17.461(011)& 17.261(011) & 18.337(139)& 17.700(014)& \ldots     & 18.038(012)& 17.461(011)& 17.257(011)\\
 41 & 218.190201 & -44.145988 & 17.464(039)& 16.742(009)& \ldots     & 17.053(008)& 16.513(007)& 16.326(007) & 17.398(039)& 16.768(009)& \ldots     & 17.056(008)& 16.513(007)& 16.322(007)\\
 42 & 218.245605 & -44.146679 & 16.749(032)& 16.019(006)& 17.404(106)& 16.334(007)& 15.811(007)& 15.623(007) & 16.683(032)& 16.044(006)& 17.372(106)& 16.337(007)& 15.811(007)& 15.619(007)\\
 43 & 218.259705 & -44.146191 & \ldots     & \ldots     & \ldots     & \ldots     & \ldots     & \ldots      & \ldots     & \ldots     & \ldots     & \ldots     & \ldots     & \ldots     \\
 44 & 218.259399 & -44.152180 & 18.892(104)& 18.123(020)& \ldots     & 18.495(018)& 17.848(013)& 17.622(014) & 18.822(104)& 18.154(020)& \ldots     & 18.498(018)& 17.848(013)& 17.617(014)\\
\enddata
\tablecomments{For convenience, table \ref{tab:LSPhot} shows the standard photometry of the local sequence stars.
The natural Swope photometry is available in the machine-readable format.}
\end{deluxetable}

\begin{deluxetable}{cccccchhh}
   \tablecaption{NIR local sequence photometry for SN2013aa\label{tab:LSNIRPhot}}
   \tablehead{
      \colhead{ID} & \colhead{$\alpha$ (2000)} & \colhead{$\delta$ (2000)} & 
      \colhead{$Y$} & \colhead{$J$} & \colhead{$H$} & \nocolhead{$j$} &
      \nocolhead{$h$}}
   \startdata
103 & 218.152359 & -44.199512 & 14.986(032)& 14.661(020)& 14.138(036)& 14.651(020)& 14.158(036)\\
104 & 218.158768 & -44.201530 & 14.610(035)& 14.191(033)& 13.492(022)& 14.178(033)& 13.519(022)\\
105 & 218.156769 & -44.231560 & 15.664(037)& 15.351(037)& 14.926(038)& 15.343(037)& 14.943(038)\\
107 & 218.161942 & -44.242981 & 12.919(030)& 12.696(023)& 12.370(028)& 12.690(023)& 12.383(028)\\
108 & 218.125824 & -44.245266 & 14.360(034)& 14.062(023)& 13.685(036)& 14.055(023)& 13.700(036)\\
109 & 218.111145 & -44.244247 & 14.173(042)& 13.866(020)& 13.488(020)& 13.859(020)& 13.503(020)\\
110 & 218.106705 & -44.232910 & 15.116(033)& 14.852(075)& 14.437(020)& 14.844(075)& 14.453(020)\\
111 & 218.119980 & -44.222977 & 12.416(032)& 12.017(031)& 11.369(116)& 12.005(031)& 11.394(116)\\
112 & 218.109863 & -44.206455 & 15.261(044)& 14.908(028)& 14.260(037)& 14.896(028)& 14.285(037)\\
113 & 218.127289 & -44.201855 & 16.274(052)& 15.992(055)& 15.752(073)& 15.987(055)& 15.761(073)\\
114 & 218.135681 & -44.211494 & 16.138(047)& 15.897(062)& 15.453(083)& 15.889(062)& 15.470(083)\\
\enddata
\tablecomments{For convenience, table \ref{tab:LSNIRPhot} shows the standard photometry of the local sequence stars.
The natural du Pont photometry for $J$ and $H$ is available in the machine-readable format.}
\end{deluxetable}

\section{Photometry of SN~2013aa and SN~2017cbv}
\label{sec:12ht+15F}
In this appendix we present the photometry for the two CSP-II objects used in 
this paper. The CSP-II was a continuation
of the \textit{Carnegie Supernova Project} \citep{Hamuy:2006}, with the goal of
obtaining NIR observations at higher red-shift on average than in the CSP-I. 
The methods for obtaining and reducing the optical and NIR photometry are detailed in
\citet{Phillips:2018}.

\begin{figure}
   \includegraphics[width=0.4\textwidth]{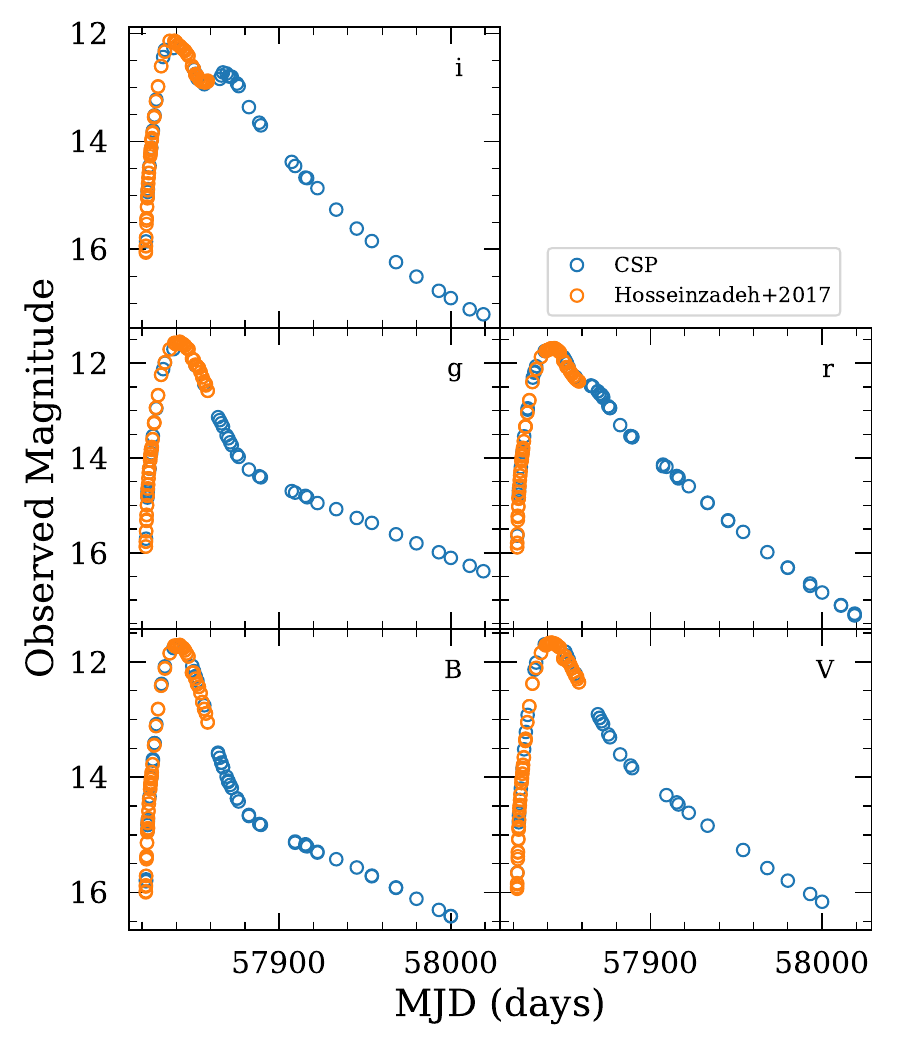}
   \caption{Comparison between the CSP photometry (blue circles) and that of 
   \citet{Hosseinzadeh:2017} (orange circles). 
   No S-corrections were applied to either dataset.\label{fig:SN2017cbv_comp}}
\end{figure}

Table~\ref{tab:Phot} lists the photometry of SN~2013aa and SN~2017cbv. 
Tables \ref{tab:LSPhot} and \ref{tab:LSNIRPhot}
list the photometry of the reference stars in the standard 
optical \citep{Landolt:1992,Smith:2002} and NIR \citep{Persson:1998} systems. Note
that there are many stars in common between the two SNe, but they were observed with
different CCDs\footnote{
SN~2013aa was observed with SITe3 and SN~2017cbv was observed with e2V. } 
and therefore have slightly different color terms (see table 5 of \cite{Phillips:2018}),
yielding different standard photometry.
The filter functions and photometric zero-points $zp_\lambda$
of the CSP-I and CSP-II
natural systems are available at the CSP website.\footnote{
\href{https://csp.obs.carnegiescience.edu}{https://csp.obs.carnegiescience.edu}}
These can be used to S-correct \citep{Stritzinger2002} the photometry to 
other systems. Figure \ref{fig:SN2017cbv_comp} shows a comparison of the CSP
photometry with that of \cite{Sand:2018}, showing very good agreement.

\section{Bayesian Hierarchical model for Intrinsic Dispersion}

\begin{deluxetable*}{llllllll}
   \tablecaption{Statistical Comparison between Surveys.
      \label{tab:stat_survey}}
\tablehead{
   \colhead{Surveys} & \colhead{$\left<\Delta \mu \right>$} & \colhead{$\sigma(sys)$} & \colhead{$\sigma(s_{BV})$} &
   \colhead{$\sigma(E(B-V))$} & \colhead{$\rho(\mu,s_{BV})$} &
   \colhead{$\rho(\mu,E(B-V))$} & \colhead{N}
}
\startdata
   CSP/LOSS   & $-0.05(03)$ & $0.16$ & $0.07$ & $0.06$ & $0.75$ & $-0.69$ & 32\\
   CSP/Swift  & $-0.02(06)$ & $0.19$ & $0.07$ & $0.06$ & $0.31$ & $-0.69$ & 11\\
   CSP/CfA    & $-0.03(01)$ & $0.09$ & $0.07$ & $0.07$ & $0.35$ & $-0.67$ & 61\\
   LOSS/Swift & $-0.07(03)$ & $0.21$ & $0.10$ & $0.09$ & $0.48$ & $-0.76$ & 26\\
   LOSS/CfA   & $0.06(02)$ & $0.16$ & $0.10$ & $0.08$ & $0.50$ & $-0.71$ & 87\\
   CFA/Swift  & $-0.07(04)$ & $0.15$ & $0.07$ & $0.08$ & $0.56$ & $-0.93$ & 19\\
\enddata
\end{deluxetable*}

In order to quantify the intrinsic dispersion in the distances to sibling
host galaxies, we construct a simple Bayesian model. For each pair of
siblings, we have the difference in the distance estimate $\Delta \mu_{i,j}$
and an associated error $\sigma(\mu_{i,j})$ which we assume to be given
by:
\begin{equation}
   \sigma(\mu_{i,j}) = \sqrt{\sigma(\mu_i)^2 + \sigma(\mu_j)^2 +
                        \sigma(sys)_{i,j}^2},
\end{equation}
\noindent
where $\sigma(\mu_i)$ and $\sigma(\mu_j)$ are the formal errors in the
distances, including photometric uncertainties and errors in the 
calibration parameters (Phillips relation, extinction, etc). We also
add additional systematic errors $\sigma(sys)_{i,j}$ when comparing 
distances using different photometric systems. These were estimated by
fitting SNe~Ia that were observed by two or more surveys and computing
the standard deviation of the differences in distance estimates. We
summarize these in Table \ref{tab:stat_survey}. We also include 
the mean difference in distance $\left<\Delta \mu \right>$, the standard
deviation in the light-curve parameters ($s_{BV}$ and $E(B-V)$),
the Pearson correlation coefficients, and
the number of SNe~Ia used. The mean offsets were not applied to
the distances in table \ref{tab:siblings} nor in the analysis to follow, 
but rather are kept as part of the error in $\sigma(sys)$.
Generally speaking, the difference in distance 
estimates are most strongly correlated with differences in estimates in the 
extinction. 

We model the true distribution of $\Delta \mu_{i,j}^T$ as a normal 
distribution centered at zero with scale $\sigma_{SN}$. The
observed $\Delta \mu_{i,j}$ are modeled as normal distributions
centered at $\Delta \mu_{i,j}^T$ with scale $\sigma(\mu_{i,j})$.
Symbolically:
\begin{eqnarray}
   \sigma_{SN} & \sim & U(0,\infty)\\
   \Delta \mu_{i,J}^T & \sim & N\left(0, \sigma_{SN}\right)\nonumber\\
   \Delta \mu_{i,j} & \sim & N\left(\Delta \mu_{i,j}^T, 
      \sigma(\mu_{i,j})\right).  \nonumber
\end{eqnarray}
We fit for the value of $\sigma_{SN}$ using Markov Chain Monte Carlo (MCMC)
methods. Table \ref{tab:sigma_SN}  lists the results with different subsets of the sibling
SNe. 
\clearpage
\bibliographystyle{aasjournal}
\bibliography{Twins.bib}

\end{document}